\documentclass[authoryear,12pt]{elsarticle}

 \usepackage{graphicx}

\usepackage{amssymb}
\usepackage{lscape}	
\usepackage{longtable}
\usepackage{rotating}
\usepackage[table]{xcolor}
\usepackage{multirow}

\journal{Icarus}

\begin{document}

\begin{frontmatter}


 \author{Sarah M. H\"orst \corref{cor1}\fnref{label1,label2}}
 \author{Y. Heidi Yoon \fnref{label2}}
 \author{Melissa S. Ugelow \fnref{label2}}
 \author{Alex H. Parker \fnref{label5}}
 \author{Rui Li \fnref{label2,label4}}
 \author{Joost A. de Gouw \fnref{label2,label4}}
 \author{Margaret A. Tolbert \fnref{label2,label3}}
 \fntext[label1]{Department of Earth and Planetary Sciences, Johns Hopkins University, Baltimore, MD, USA}
 \fntext[label2]{Cooperative Institute for Research in Environmental Sciences, University of Colorado, Boulder, CO, USA}
 \cortext[cor1]{sarah.horst@jhu.edu}
 \fntext[label5]{Southwest Research Institute, Department of Space Studies, Boulder, CO, USA}
 \fntext[label4]{Chemical Sciences Division, National Oceanic and Atmospheric Administration, Earth System Research Laboratory, Boulder, CO, USA}
  \fntext[label3]{Department of Chemistry and Biochemistry, University of Colorado, Boulder, CO, USA}

\title{Laboratory Investigations of Titan Haze Formation: In Situ Measurement of Gas and Particle Composition}




\begin{abstract}
Prior to the arrival of the Cassini-Huygens spacecraft, aerosol production in Titan's atmosphere was believed to begin in the stratosphere where chemical processes are predominantly initiated by far ultraviolet (FUV) radiation. However, measurements taken by the Cassini Ultraviolet Imaging Spectrograph (UVIS) and Cassini Plasma Spectrometer (CAPS) indicate that haze formation initiates in the thermosphere where there is a greater flux of extreme ultraviolet (EUV) photons and energetic particles available to initiate chemical reactions, including the destruction of N$_{2}$. The discovery of previously unpredicted nitrogen species in measurements of Titan's atmosphere by the Cassini Ion and Neutral Mass Spectrometer (INMS) indicates that nitrogen participates in the chemistry to a much greater extent than was appreciated before Cassini. The degree of nitrogen incorporation in the haze particles is important for understanding the diversity of molecules that may be present in Titan's atmosphere and on its surface. We have conducted a series of Titan atmosphere simulation experiments using either spark discharge (Tesla coil) or FUV photons (deuterium lamp) to initiate chemistry in CH$_{4}$/N$_{2}$ gas mixtures ranging from 0.01\% CH$_{4}$/99.99\% N$_{2}$ to 10\% CH$_{4}$/90\% N$_{2}$. We obtained \emph{in situ} real-time measurements using a high-resolution time-of-flight aerosol mass spectrometer (HR-ToF-AMS) to measure the particle composition as a function of particle size and a proton-transfer ion-trap mass spectrometer (PIT-MS) to measure the composition of gas phase products. These two techniques allow us to investigate the effect of energy source and initial CH$_{4}$ concentration on the degree of nitrogen incorporation in both the gas and solid phase products. The results presented here confirm that FUV photons produce not only solid phase nitrogen bearing products but also gas phase nitrogen species. We find that in both the gas and solid phase, nitrogen is found in nitriles rather than amines and that both the gas phase and solid phase products are composed primarily of molecules with a low degree of aromaticity. The UV experiments reproduce the absolute abundances measured in Titan's stratosphere for a number of gas phase species including C$_{4}$H$_{2}$, C$_{6}$H$_{6}$, HCN, CH$_{3}$CN, HC$_{3}$N, and C$_{2}$H$_{5}$CN.
\end{abstract}


\end{frontmatter}

\newpage

\section*{Highlights}
\begin{itemize}
\item Heaviest gas phase products detected from Titan atmosphere simulation experiment
\item Demonstrate presence of nitrogen in the gas- and condensed- phase products using FUV photons
\item UV experiments reproduce abundances of gases measured in Titan's stratosphere 
\item Nitrogen found in nitriles in both gas and solid phase for UV and spark
\item Investigate Titan haze formation using different energy sources with other variables constant
\end{itemize}


\section{Introduction} 

The discovery of very heavy ions (negative ions with mass-to-charge up to 10,000 amu/q \citep{Coates:2007}, positive ions with mass-to-charge up to 400 amu/q \citep{Crary:2009}) in Titan's thermosphere has dramatically altered our understanding of the processes involved in the formation of the complex organic aerosols that comprise Titan's characteristic haze. Prior to the arrival of the Cassini-Huygens spacecraft, aerosol production was believed to begin in the stratosphere where chemical processes are predominantly initiated by far ultraviolet (FUV) radiation. However, the discovery of very heavy ions, coupled with Cassini Ultraviolet Imaging Spectrograph (UVIS) occultation measurements that show haze absorption up to 1000 km \citep{Liang:2007}, indicates that haze formation initiates in the thermosphere. The energy environment of the thermosphere is significantly different from the stratosphere; in particular there is a greater flux of extreme ultraviolet (EUV) photons and energetic particles available to initiate chemical reactions, including the destruction of N$_{2}$, in the upper atmosphere. The discovery of previously unpredicted nitrogen species in measurements of Titan's atmosphere by the Cassini Ion and Neutral Mass Spectrometer (INMS) indicates that nitrogen participates in the chemistry to a much greater extent than was appreciated before Cassini \citep{Vuitton:2007}. Additionally, measurements obtained by the Aerosol Collector Pyrolyzer (ACP) carried by Huygens to Titan's surface may indicate that Titan's aerosols contain significant amounts of nitrogen \citep{Israel:2005, Israel:2006}. Additionally, \citet{Sebree:2014} recently showed that the inclusion of nitrogen bearing aromatic precursors in laboratory atmosphere simulation experiments results in particles that are a better fit to the far IR spectrum of Titan aerosols obtained from Cassini CIRS. Previous Titan atmosphere simulation experiments have also demonstrated that the presence of nitrogen increases the complexity of both the gas and aerosol phase chemistry \citep{Fujii:1999, Imanaka:2007, Imanaka:2009, Trainer:2012} and may play a key role in increasing the efficiency of gas to particle conversion \citep{Trainer:2012}.The degree of nitrogen incorporation in the haze particles is important for understanding the diversity of molecules that may be present in Titan's atmosphere and on its surface. The building blocks of life as we know it, such as amino acids and nucleotide bases, require the presence of nitrogen and previous Titan atmosphere simulation experiments have demonstrated that gas phase and/or heterogeneous chemistry results in the formation of some of these building blocks including molecules such as adenine (C$_{5}$N$_{5}$H$_{5}$) \citep{Horst:2012}. 

For nearly 50 years, mixtures of N$_{2}$ and CH$_{4}$ have been subjected to a variety of energy sources in numerous different experimental setups to produce Titan aerosol analogues, so-called ``tholins'', and investigate both the gas and solid phase products. These energy sources include protons, laser induced plasma (LIP), gamma rays, corona discharge, electrical discharge, UV lamps (Hg or deuterium), and UV synchrotron. Recent comprehensive reviews of Titan atmosphere simulation experiments, including a detailed discussion of various energy sources, can be found in \citet{Cable:2012} and \citet{Coll:2013}. It can be very difficult to compare the results of simulation experiments from different laboratories because of the differences in experimental setup (temperature, pressure, gas mixture, and energy source) and analysis techniques. Our unique experimental setup allows for direct comparison between the products formed by two different energy sources, spark discharge and UV, without any other differences in experimental technique, materials, or analysis techniques. This allows for investigation of the effect of energy source on the incorporation of nitrogen into tholin without any other variables. 

We have conducted a series of Titan atmosphere simulation experiments using either spark discharge (Tesla coil) or FUV photons (deuterium lamp) to initiate chemistry in CH$_{4}$/N$_{2}$ gas mixtures ranging from 0.01\% CH$_{4}$/99.99\% N$_{2}$ to 10\% CH$_{4}$/90\% N$_{2}$. We have obtained \emph{in situ} measurements using three different techniques. We use a high-resolution time-of-flight aerosol mass spectrometer (HR-ToF-AMS) to measure the particle composition as a function of particle size, a proton-transfer ion-trap mass spectrometer (PIT-MS) to measure the composition of gas phase products, and a scanning mobility particle sizer (SMPS) to measure particle size distributions. These three techniques allow us to investigate the effect of energy source and initial CH$_{4}$ concentration on the degree of nitrogen incorporation in both the gas and solid phase products, the size and number of particles, and particle density. The size and density measurements from these experiments have been published in \citet{Horst:2013} and will be discussed here only as they relate to interpretation of the other measurements. 

\section{Materials and Experimental Methods}
\subsection{Haze Production setup}

\begin{figure}
\centering
\resizebox{5.5in}{!}
{\includegraphics{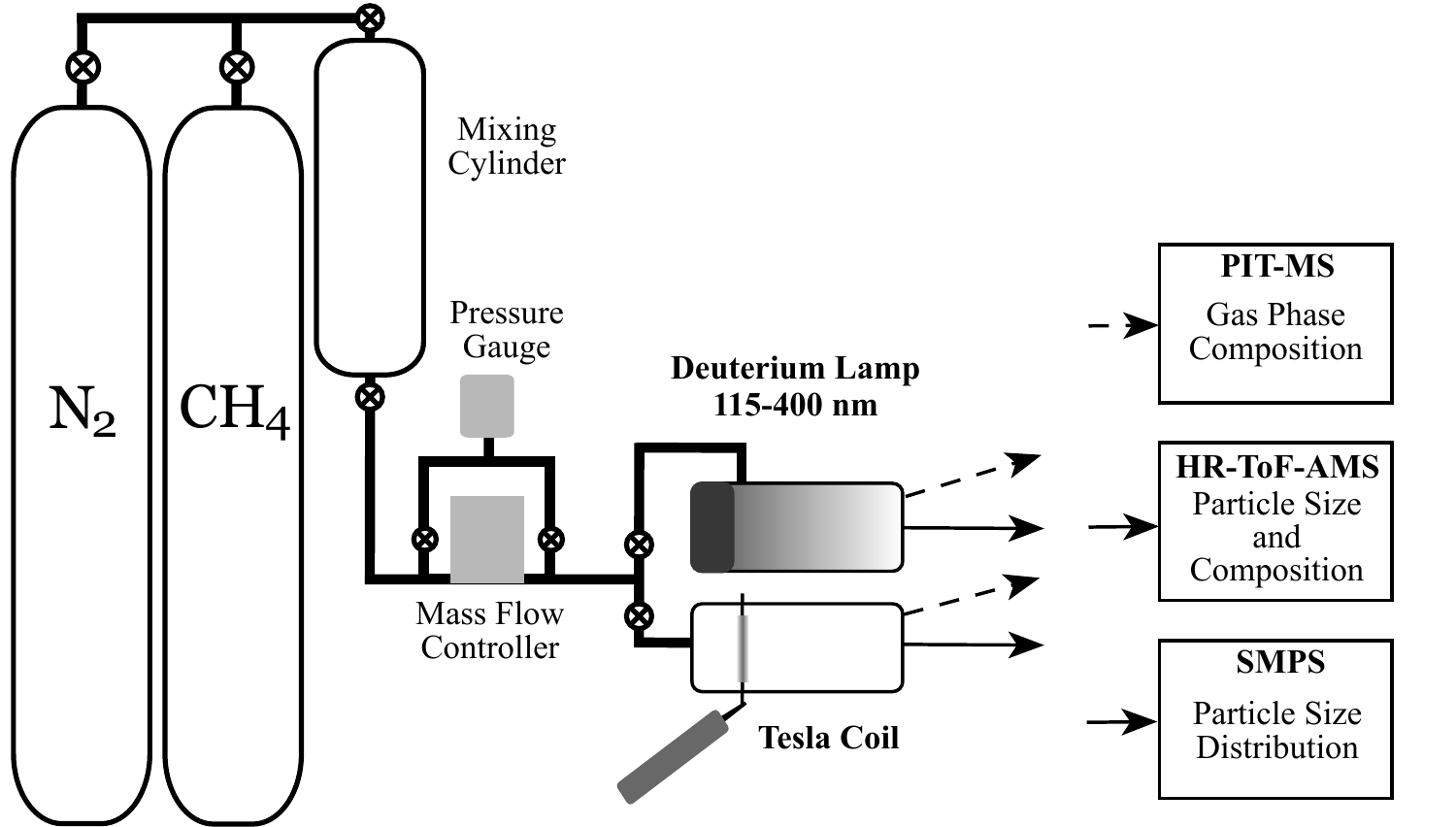}}
\caption{Schematic of the experimental setup used for this work. N$_{2}$ and CH$_{4}$ are mixed overnight in the mixing cylinder. Gases flow through one of two reaction cells (UV or spark) where they are exposed to FUV photons from a deuterium lamp or the electric discharge produced by a Tesla coil initiating chemical processes that lead to the formation of new gas phase products and particles. The composition of the  gas phase products is then measured using a proton-transfer ion-trap mass spectrometer (PIT-MS). The particles are analyzed using either a high resolution time-of-flight aerosol mass spectrometer (HR-ToF-AMS) to measure their composition or a scanning mobility particle sizer (SMPS) to measure their size distribution. Each reaction cell has two outlets; the dashed lines indicate where teflon lines, fittings, and valves were used while the solid lines indicate stainless steel lines, fittings, and valves. All experiments were run at room temperature and 620-640 Torr (atmospheric pressure in Boulder, Colorado). \label{fig:experiment}}
\end{figure}

A schematic of the experimental setup is shown in Figure \ref{fig:experiment}. Previous spark experiments were performed using a similar setup by \citet{Trainer:2004b, Trainer:2004} and \citet{Horst:2013} and previous UV experiments were performed by \citet{Trainer:2006, Trainer:2012, Trainer:2013} and \citet{Horst:2013}. Reactant gases, N$_{2}$ (99.999\% Airgas) and CH$_{4}$ (99.99\% Airgas), are mixed in a stainless steel mixing chamber. CH$_{4}$ was introduced in mixing ratios ranging from 0.01\% to 10\% (see Table \ref{table:exp}) and then the mixing chamber was filled to 600 PSI with N$_{2}$ and allowed to mix for a minimum of 8 hours. The reactant gases were continuously flowed through a glass reaction cell using a mass flow controller (Mykrolis FC-2900). Four flow rates were investigated, low ($\sim$60 sccm), standard ($\sim$100 sccm), medium ($\sim$130 sccm), and high ($\sim$230 sccm). The pressure in the reaction cell was measured (MKS Baratron) and maintained between 620 and 640 Torr at room temperature. The reactant gases were exposed to one of two energy sources, spark discharge or UV. The setup is identical for both types of experiments until the gases reach the glass reaction cells. The spark reaction cell has two electrodes, one connected to a Tesla coil (Electro Technic Products) and the other connected to ground, with a separation of 3.8 cm. The UV reaction cell is connected to a deuterium continuum lamp (115-400 nm, Hamamatsu L1835, MgF$_{2}$ window). Both the spark cell and the UV cell have two outlets, which allowed for simultaneous measurements by different instruments, or facilitated switching between instruments without stopping particle production. 

\begin{table}
\begin{center}
\caption{Summary of Experiments\label{table:exp}}
\begin{tabular}{l|lllll|llll}
\hline

&\multicolumn{5}{c}{UV}&\multicolumn{4}{c}{Spark}\\ 
\raisebox{1.5ex}[0pt]{\hspace{0.3cm} [CH$_{4}$]}&&\multicolumn{8}{c}{Flow Rate (sccm)}\\ \cline{3-10}
\raisebox{1.3ex}[0pt]{(Initial \%)}&$\tau$$^{*}$&60&100&130&230&60&100&130&230\\
\hline
0.01&0.81&&A,S&&A&&A,S&&\\
0.1&8.1&A,S&A,S&A,P&A,P&A,S&A,S&A,S,P&A,S,P\\
1&81&A,S&A,S&A,S&A,S,P&A,S&A,S&A,S&A,S\\
2&160&A,S&A,S&A,S,P&A,S,P&A,S&A,S&A,S,P&A,S,P\\
5&405&&A,S&&&A,S&A,S&A,S&A,S,P\\
10&810&&A,S&P&&A,S&A,S&A,S,P&A,S,P\\
\hline
\multicolumn{10}{l}{}\\
\multicolumn{10}{l}{AMS (A), SMPS (S), and P (PIT-MS) performed at a variety of flow}\\
\multicolumn{10}{l}{rates and CH$_{4}$ mixing ratios. PIT-MS measurements could not be}\\
\multicolumn{10}{l}{obtained for the 60 and 100 sccm flow rates because of instrument}\\
\multicolumn{10}{l}{constraints. AMS and SMPS measurements were repeated at least}\\
\multicolumn{10}{l}{twice, while the PIT-MS measurements were only performed once due}\\
\multicolumn{10}{l}{to limited instrument access. PIT-MS and AMS measurements were}\\
\multicolumn{10}{l}{obtained simultaneously for the 230 sccm experiments.}\\
\multicolumn{10}{l}{$^{*}$ Approximate Lyman-$\alpha$ optical depth ($\tau$) at the end of the UV reaction cell}\\
\end{tabular}
\end{center}
\end{table}

\begin{table}
\begin{center}
\caption{Residence time (minutes) of gases in energetic regime\label{table:res}}
\vspace{0.3cm}

\begin{tabular}{l|lllllll}
\hline

&\multicolumn{6}{c}{UV}&Spark\\ 
\raisebox{1.5ex}[0pt]{\hspace{0.2cm} Flow Rate}&\multicolumn{6}{c}{[CH$_{4}$] (Initial \%)}&\\ \cline{2-7}
\raisebox{1.3ex}[0pt]{\hspace{0.6cm}(sccm)}	&	0.01	&	0.1	&	1	&	2	&	5	&	10	&\\	
\hline
60	&	6.9	&	0.69	&	0.069	&	0.034	&	0.014	&	0.0069	&	0.010\\
100	&	4.6	&	0.46	&	0.046	&	0.023	&	0.0092	&	0.0046	&	0.007\\
130	&	3.1	&	0.31	&	0.031	&	0.015	&	0.0062	&	0.0031	&	0.005\\
230	&	2.4	&	0.24	&	0.024	&	0.012	&	0.0045	&	0.0024	&	0.003\\

\hline
\multicolumn{8}{l}{For the FUV lamp, the residence time is the time before reaching}\\
\multicolumn{8}{l}{the region of the cell where CH$_{4}$ is optically thick ($\tau$=1)}.\\
\multicolumn{8}{l}{For the Tesla coil, the region of exposure to the plasma does not }\\
\multicolumn{8}{l}{vary as a function of CH$_{4}$ concentration.}\\
\end{tabular}
\end{center}
\end{table}

Aerosol, or tholin, is produced through chemistry initiated by the energy from the Tesla coil or deuterium lamp. A thorough description of the various energy sources used to initiate tholin formation, including their strengths and weaknesses, is found in \citet{Cable:2012}. Briefly, we use the electrical discharge from the Tesla coil because it is an energy source that is known to dissociate N$_{2}$. Although the energy density is higher than the energy available to initiate chemistry in Titan's atmosphere, we use it as an analogue for the relatively energetic environment of Titan's upper atmosphere. The Tesla coil operates at a range of voltages and for these experiments, the Tesla coil was set to optimize the amount of tholin produced using 2\% CH$_{4}$ for our analytical techniques. The same setting was used for all other experiments and was periodically checked using 2\% CH$_{4}$ to ensure that the production rate had not changed. Using a calibration device from the manufacturer of the Tesla coil, the voltage dissipated in the cell was measured to be 20-25 kV, with a current of 0.1 mA, resulting in a power of 2-2.5 watts. Note that this is an upper limit as some power that is lost may not be dissipated in the gases themselves. Previous work using this experimental setup demonstrated that variation in the initial gas mixture composition did not measurably affect the discharge power \citep{Trainer:2004}.

\begin{figure}
\centering
\resizebox{5.5in}{!}
{\includegraphics{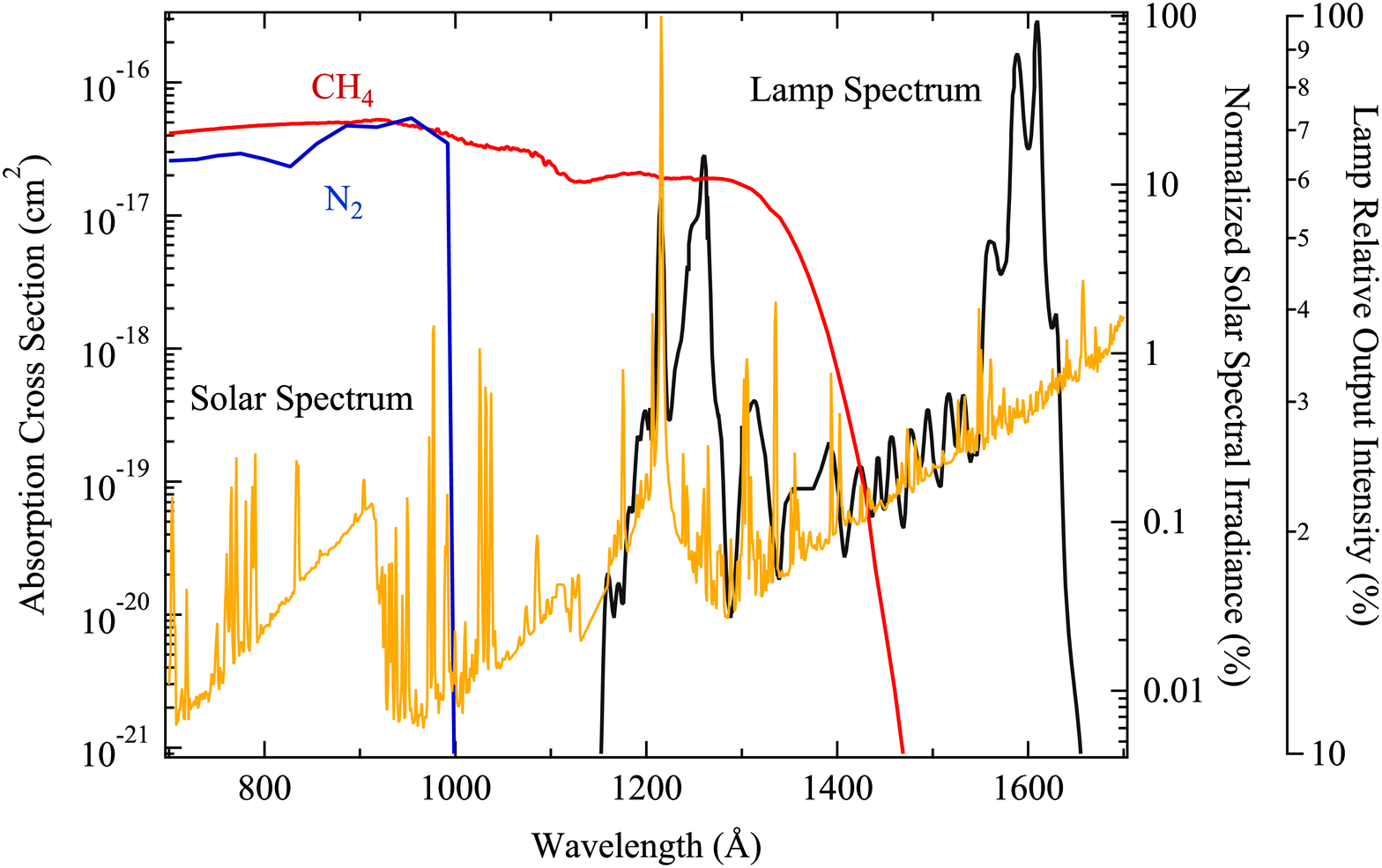}}
\caption{Comparison of the absorption cross sections of N$_{2}$ (blue) \citep{Chan:1993d} and CH$_{4}$ (red) (left axis) \citep{Chen:2004, Kameta:2002, Mount:1977} to the solar spectrum (orange) (right, interior axis) \citep{Woods:2009} and the spectrum of the deuterium lamp used in these experiments (black) (right, exterior axis) (as provided by the manufacturer, Hamamatsu). Note that the N$_{2}$ cross section is highly structured, which is not captured in this simplified representation. \label{fig:photo_ch4}}
\end{figure}

Although energetic electrons and particles do contribute to the chemistry occurring in Titan's atmosphere, photons are the dominant source of energy for ionization and dissociation \citep{Lavvas:2011}. We are therefore also interested in exploring these processes. While FUV photons produced by the deuterium lamp used in this experiment (115-400 nm, main peaks near 121.6 nm, 125.4 nm, and 160.8 nm) are not sufficiently energetic to directly dissociate N$_{2}$, we do dissociate CH$_{4}$. The relevant absorption cross-sections are shown in Figure \ref{fig:photo_ch4}. However, \citet{Trainer:2012} demonstrated that nitrogen is indeed participating in the chemistry in our reaction cell, most likely through the reaction of N$_{2}$ with CH (ground or excited state) formed from CH$_{4}$ dissociation. However, more information about the rates and branching ratios of reactions is required before the mechanism can be determined. A detailed discussion of a number of possibilities can be found in \citet{Trainer:2012} and \citet{Yoon:2014}. The results presented here confirm that nitrogen bearing products are not only present in the tholins, but are also present in the gas phase. \citet{Trainer:2006} used N$_{2}$O actinometry to measure the photon flux of the lamp and found a flux of $4.4\times10^{15}$ photons/s across the wavelength range of the lamp, which corresponds to $\sim7\times10^{-3}$ watts assuming the energy of a lyman-alpha photon. N$_{2}$O does not absorb photons covering the full wavelength range of the lamp (see e.g., \citet{Rajappan:2010}); thus the measurements of \citet{Trainer:2006} underestimate the photon flux and therefore energy deposition from photons in the experiment by approximately a factor of 2 based on the lamp spectrum provided by Hamamatsu. The lamp energy deposition rate used here is accordingly a lower limit.

Although the energy deposition rate is lower for the FUV lamp than for the Tesla Coil, as shown in Table \ref{table:res}, the gases are exposed to the energy from the Tesla Coil for a shorter period of time. Combining the deposition rates and residences times we can estimate the total amount of energy; for the 0.01\% CH$_{4}$ experiments at all flow rates the total energy deposited is higher for the UV experiments than for the spark experiments, while the higher concentration CH$_{4}$ experiments are either comparable (for the 0.1\% CH$_{4}$ experiment) or have lower total energy deposition (for the 1\%, 2\%, 5\%, 10\% CH$_{4}$ experiments). We note again that the FUV energy deposition rate is a lower limit, while the Tesla Coil energy deposition rate is an upper limit. If the total energy deposited is the most important parameter determining the chemistry occurring in these experiments, we would expect the results from the Tesla Coil experiments to fall between the results for the 0.01\% and 0.1\% CH$_{4}$ UV experiments; as shown below, this is not the case. However, the total energy deposited for the UV experiments does appear to affect the particle production rate as discussed in Section 2.5.


The gas and particle phase products produced in the reaction cell flow out of the cell and into one of three instruments, a high resolution time-of-flight aerosol mass spectrometer (HR-ToF-AMS), a scanning mobility particle sizer (SMPS) or a proton-transfer ion-trap mass spectrometer (PIT-MS). The HR-ToF-AMS and SMPS measure particle phase products, while the PIT-MS measures gas phase products. Stainless steel (304 or 316) fittings, tubing, and valves were used to connect the reaction cell to the HR-ToF-AMS and SMPS to maximize particle transmission, while teflon (PFA) fittings, tubing, and valves were used to connect the reaction cell to the PIT-MS to maximize gas phase species transmission. AMS and SMPS measurements were repeated at least twice, while the PIT-MS measurement were only performed once due to limited instrument access. Table \ref{table:exp} contains a summary of the experiments we performed.

One advantage of the three \emph{in situ} techniques is that all measurements are obtained without exposing the products to Earth's atmosphere and without allowing the samples to sit in the laboratory and age. However, as with most tholin experiments (see e.g., \citet{Cable:2012}) we see evidence of oxygen contamination in both the gas and particle phase products of these experiments. The particles are generally ~5-10\% oxygen by mass, but HR-ToF-AMS measurements indicate a large fraction of that is adsorbed water rather than oxygen incorporated by chemistry. This is consistent with a number of other experiments \citep{Cable:2012}. One disadvantage of these techniques is that they necessitate particle production rates high enough to produce sufficient aerosol for real time analysis; for this reason we typically run our experiments at 620-640 Torr (atmospheric pressure in Boulder, Colorado, altitude $\sim$1600 m). These pressures are higher than the pressures in the regions of Titan's atmosphere where the chemistry leading to aerosol formation begins. For that reason, work is ongoing in our laboratory to investigate any pressure dependence in our results and to ascertain the range of pressures for which experiments can be run based on our experimental and analytical constraints. Results of preliminary pressure dependent studies involving benzene photolysis can be found in \citet{Yoon:2014}. For the purposes of this work, we are interested in comparing differences resulting only from choice of energy source and CH$_{4}$ concentration at our standard experimental pressure. 

\subsection{High Resolution Time-of-Flight Mass Spectrometry (HR-ToF-AMS)}

\begin{figure} 
\centering
\resizebox{5.5in}{!}
{\includegraphics{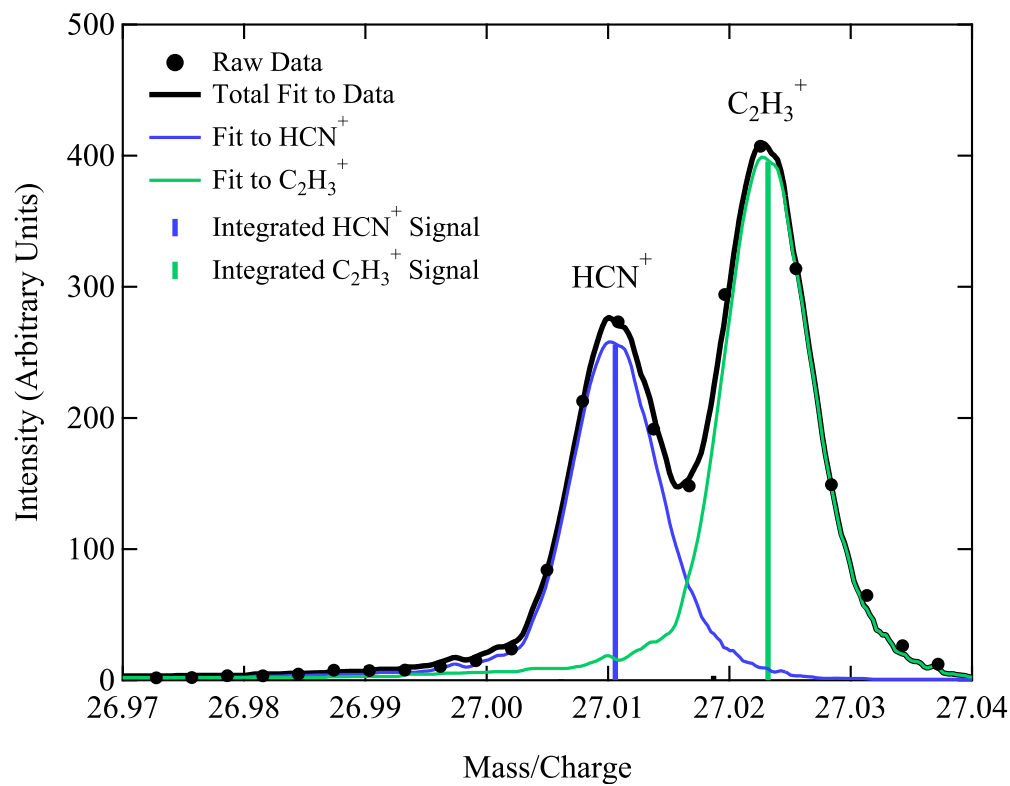}}
\caption{The resolution of the HR-ToF-AMS is sufficiently high to allow for differentiation between nitrogen bearing species and hydrocarbon species. Shown here is a comparison between the raw HR-ToF-AMS data (circles), the individual peak fits for HCN$^{+}$ and C$_{2}$H$_{3}^{+}$ (blue and green lines, respectively), the total fit at mass/charge 27 (black line) and the integrated signal for HCN$^{+}$ and C$_{2}$H$_{3}^{+}$ for typical tholin (blue and green sticks, respectively). \label{fig:resolution}}
\end{figure}

We use a high-resolution time-of-flight aerosol mass spectrometer (HR-ToF-AMS, or AMS, Aerodyne Research) to measure the particle composition and particle size. Details of the instrument design can be found in \citet{decarlo:2006} and use of the AMS with this experimental setup has been described previously in \citet{Trainer:2012, Trainer:2013} and \citet{Horst:2013}. Briefly, the particles exit the reaction chamber and flow through a critical orifice (120 $\mu$m, SPI, Inc.) and are then focused by an aerodynamic lens, which transmits particles with aerodynamic diameters ($D_{va}$) ranging from $\sim$20 nm to $\sim$1 $\mu$m into the particle time-of-flight region. As shown in \citet{Horst:2013}, the size range is sufficient to include all of the particles produced in the experiment. The particles are flash vaporized on a resistively heated surface at a temperature of approximately 600$^{\circ}$C. The resulting molecules are then ionized via 70 eV electron ionization. The ions are analyzed by a high-resolution time-of-flight mass spectrometer (H-TOF Platform, Tofwerk, Thun, Switzerland). The data presented here were collected using the ``W-mode" of the HR-ToF-AMS, which has a longer flight path than the ``V-mode" resulting in better mass resolution (M/$\Delta$M $\sim$ 3000-4300 for \emph{m/z} $<$200) but lower sensitivity \citep{decarlo:2006}. This resolution is sufficient to investigate nitrogen incorporation in the aerosol, as shown in Figure \ref{fig:resolution}, which demonstrates our ability to resolve HCN$^{+}$ and C$_{2}$H$_{3}^{+}$ at \emph{m/z} 27.

Measurements were also obtained using the particle time-of-flight (PToF) mode of the instrument. In PToF mode, the size dependent velocities of the particles in the particle time-of-flight region, obtained by chopping the particle beam to allow for arrival time measurements, are used to determine the particle vacuum aerodynamic diameter ($D_{va}$) \citep{Jimenez:2003a, Jimenez:2003b, decarlo:2006}. $D_{va}$ is defined as the diameter of a unit density sphere that reaches the same terminal velocity in the AMS as the measured particle. In this work, we use the PToF mode to differentiate between gas phase and particle phase signals, which aids in removal of any gas phase signal from the aerosol composition measurements. $D_{va}$, in combination with $D_{m}$ determined by SMPS measurements, also allows for particle density calculations, as described in \citet{Horst:2013}. 

The AMS data were analyzed using a combination of the custom IGOR AMS analysis software programs SQUIRREL and PIKA \citep{decarlo:2006, Aiken:2007, Aiken:2008} and custom IDL software written at the University of Colorado. The mass/charge calibration of the data is performed using known ion peaks both during acquisition and during analysis. Although some peaks correspond to the signal from only a single ion species, most peaks result from overlapping peaks of overlapping ion species (see Figure \ref{fig:resolution}). Peaks are fit using an array of ions determined by the user. To quantify the signal in each peak, a modified gaussian shape and width, which are functions of the ion mass, are used. These functions are determined by fitting known peaks. The total ion signal is determined from the area of the signal peak, not the height of the peak \citep{decarlo:2006}. The total ion signal is then converted to mass concentration using ionization efficiency calibrations performed on the instrument \citep{Drewnick:2005}. Although the AMS significantly increases the ratio of particle to gas species before they enter the mass analyzer through the use of aerodynamic lenses and differential pumping, gas phase species are still present in the spectra. The gas phase signal is subtracted through the use of background spectra, which are obtained by diverting the flow through a filter (Pall Corp. P/N 12144) to remove all particles before the gas enters the AMS. The removal of gas phase signal is also checked through the use of the PToF measurements, which can easily differentiate gas phase species from larger particle species. 

Since ions are formed via electron ionization (EI), the parent molecules in the particles are fragmented and these fragments are the ions observed in the AMS measurements. It is therefore difficult to identify the parent molecules, but extensive previous calibration work has shown that elemental analysis is possible \citep{Aiken:2007, Aiken:2008, Kroll:2011, Chhabra:2010}. We do not fit ions past mass/charge 100, even though they are often present, because the resolution of the instrument is not sufficient to confidently identify ions produced in our experiments. However, this does not have a significant effect on our overall results since the fragmentation resulting from EI breaks higher mass parent molecules into lower mass, identifiable, ions. 

\subsection{Proton-Transfer Ion-Trap Mass Spectrometry (PIT-MS) \label{sect:pitms}}

To characterize the composition of the gas phase products, we use a proton-transfer ion-trap mass spectrometer (PIT-MS). PIT-MS provides real-time, time-resolved, quantitative measurements of gas phase species and does not require use of a collection technique, such as a cold trap, which might alter the results. A detailed description of the instrument can be found in \citet{deGouw:2007} and \citet{Warneke:2005}. Briefly, the gases from reaction chamber flow directly into the drift tube reactor of the PIT-MS, with an inlet flow rate of 115 sccm (about 15 sccm go into the drift tube, the rest are pumped away). There the gases encounter H$_{3}$O$^{+}$ ions from a hollow-cathode ion source. Trace gases with a higher proton affinity than H$_{2}$O are then ionized via proton-transfer with H$_{3}$O$^{+}$ (H$_{3}$O$^{+}$+R$ \rightarrow $RH$^{+}$+H$_{2}$O). This reaction is exothermic if the proton affinity of the trace gas is higher than the proton affinity of water. This technique is advantageous because it is a soft ionization source, unlike the electron ionization used in the AMS, which prevents fragmentation of the molecules. This lack of fragmentation differentiates the present study from all previous \emph{in situ} neutral gas phase measurements from these types of experiments. However, this means that a number of the molecules that form in Titan's atmosphere, and presumably form in our experiment, cannot be detected with this technique. In particular, small alkanes ($<$C$_{7}$), alkenes ($<$C$_{3}$), and alkynes ($<$C$_{3}$) cannot be detected. However, all of the previously detected nitrogen bearing species produced by Titan photochemistry can be measured by this technique, as can benzene and other aromatic species \citep{Hunter:1998}. Once ionized, an electric field within the reactor limits cluster-ion formation and transports the ions the length of the tube where ion lenses then extract the ions into the ion trap. The mass spectrometer has unit resolution over a mass range of 18-217 amu/q with a detection limit of 10-110 pptv \citep{deGouw:2007, Warneke:2005}.

The ion signals are normalized to a standard H$_{3}$O$^{+}$ signal to allow for determination of product ion abundances at a constant primary ion signal. We obtain background spectra through the use of a catalytic converter to remove volatile organic compounds. The instrument alternates between background measurements and signal measurements every few minutes. We then calculate the absolute mixing ratios from the background subtracted spectra using previously measured calibration factors. A number of species such as acetonitrile (CH$_{3}$CN) and benzene (C$_{6}$H$_{6}$) have measured calibration factors, while a standard calibration factor is used for most other masses \citep{deGouw:2007, Warneke:2005, Warneke:2011, Warneke:2011b}. HCN has a proton affinity that is close to that of water; therefore we cannot neglect the backward reaction between protonated HCN and water. Formaldehyde is a similar case and the sensitivity is reduced by a factor of 3-4 compared to the standard calibration value \citep{Warneke:2011b}. We therefore reduce the calibration value by a factor of 3.5 for HCN and increase the value of the uncertainty in the calibration accuracy. 

We averaged a minimum of 300 spectra obtained after the signals stabilized to calculate the values reported here. NO$^{+}$ (mass 30) and O$_{2}^{+}$ (mass 32) are known impurities in the instrument. Their abundances are stable on the timescale of an experiment and are therefore removed when the spectra are background subtracted. 

\subsection{Scanning Mobility Particle Sizer (SMPS)}

As described in \citet{Horst:2013}, we used a scanning mobility particle sizer (SMPS) to measure the particle size distribution \emph{in situ}. The SMPS consists of a electrostatic classifier (TSI Model 3080) with a differential mobility analyzer (DMA, TSI Model 3081) and a condensation particle counter (CPC, TSI Model 3775). The polydisperse aerosol enters the DMA, which applies an electric field to the flow of particles and then size selects them based on their electrical mobility against the drag force provided by the sheath flow. The size-selected particles then enter the CPC, which measures the number of particles by light scattering. Thus the SMPS provides the number of particles as a function of their mobility diameter ($D_{m}$). We used sheath flows of either 3 L/min or 10 L/min, corresponding to particle diameter ranges of 14.5 to 673 nm or 7.4 to 289 nm, respectively. Due to experimental constraints on flow rate, SMPS measurements were not obtained simultaneously with either PIT-MS or AMS measurements. Rather, measurements were obtained consecutively by switching from the SMPS to the AMS once particle production stabilized and a sufficient number of SMPS measurements were obtained. The switch did not necessitate stopping particle production because the reactions cells have two outlets as shown in Figure \ref{fig:experiment}. 

\subsection{Flow Rate}

The AMS, PIT-MS, and SMPS all have different instrumental constraints on experimental flow rate. As described above we investigated four flow rates, low ($\sim$60 sccm), standard ($\sim$100 sccm), medium ($\sim$130 sccm), and high ($\sim$230 sccm). These flow rates were chosen based on the optimal operating conditions for the various instruments and the desire to conduct simultaneous gas and particle phase measurements, which required approximately twice the standard flow rate. Whenever possible, multiple flow rates were tested to ensure that the difference in optimal flow rate between the instruments was not significantly affecting the results. 

The AMS requires a flow rate of $\sim$100 sccm, or the ``standard'' flow rate for these experiments. For experiments conducted at higher flow rates (``medium" and ``high") the additional flow was either sent into another instrument, such as the PIT-MS, or was removed through an additional line placed immediately before the AMS inlet. For experiments conducted at the ``low" flow rate, an additional flow of ultra high purity nitrogen was added immediately before the AMS instrument inlet. For the high flow rate experiments that included both the AMS and PIT-MS, see Table \ref{table:exp}, the measurements were obtained simultaneously. Due to experimental constraints, the ``medium" flow rate measurements were obtained separately, but consecutively. Standard and low flow measurements could not be performed using the PIT-MS because those flow rates are below the minimum instrument requirement.

The SMPS requires a higher flow rate than the AMS or PIT-MS, we therefore add an additional flow of N$_{2}$ after the particles exit the reaction chamber to bring the total flow rate to 260 sccm. The measurements presented here are corrected for the dilution caused by the additional flow of N$_{2}$.

\begin{figure}
\centering
\resizebox{5.5in}{!}
{\includegraphics{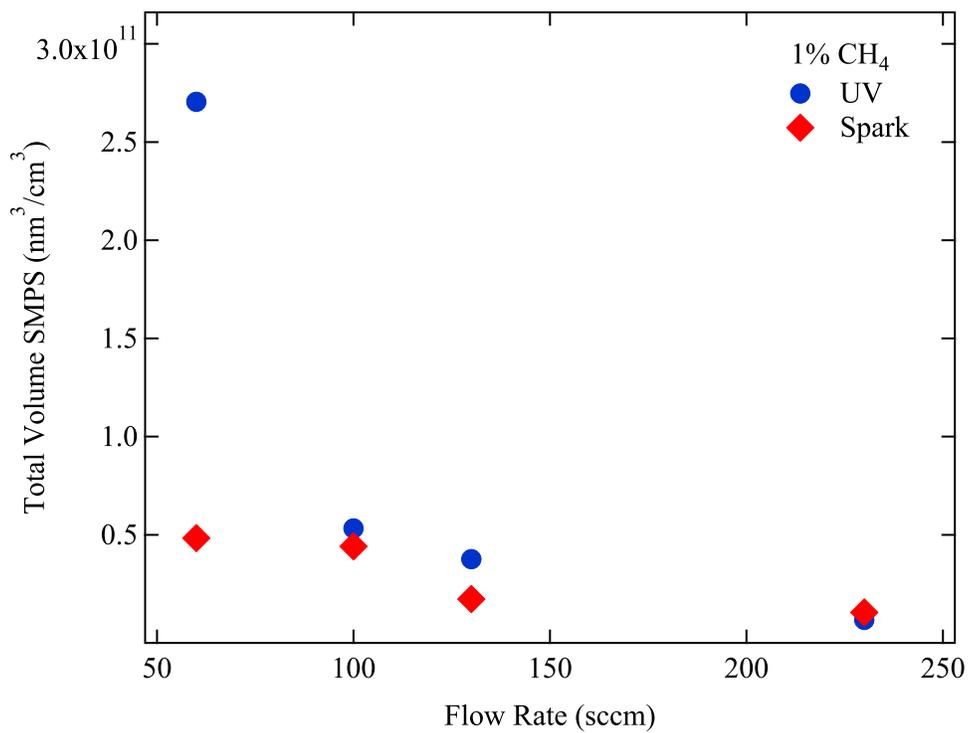}}
\caption{Flow rate has a stronger effect on the amount of aerosol produced for UV experiments, than for spark experiments. Shown here are SMPS measurements of the total volume of aerosol produced as a function of flow rate for aerosols produced from 1\% CH$_{4}$/99\% N$_{2}$ using UV (blue circles) or spark (red diamonds) energy source.\label{fig:flow}}
\end{figure}

Figure \ref{fig:flow} shows SMPS measurements of total volume of aerosol produced from 1\% CH$_{4}$ using either spark or UV for the four flow rates investigated. The total volume of aerosol produced decreases by approximately an order of magnitude from the low flow rate to the high flow rate for the UV experiments, while it decreases by about a factor of 3 for the spark experiments. As discussed in \citet{Horst:2013}, optical depth likely plays an important role in UV experiments. The optical depth in our reaction cell increases with increasing distance from the lamp (due to the increasing CH$_{4}$ column encountered) and is a function of initial CH$_{4}$ concentration and wavelength. For a given gas mixture, the amount of time the gases and particles spend in the optically thin region of the cell ($T_{\tau <1}$) is flow rate dependent. At CH$_{4}$ concentrations less than ~0.02\%, the entire reaction cell is optically thin at Lyman-$\alpha$ but since the flow rate determines the amount of time gases spend in the reaction cell, $T_{\tau <1}$ is still flow rate dependent. Figure \ref{fig:flow} shows the measured SMPS total aerosol volumes as a function of flow rate from 60 sccm to 230 sccm for experiments using 1\% CH$_{4}$. It is clear that the aerosol production rate is much more strongly affected by the change in flow rate when using the UV energy source than for the spark, where the plasma interaction region is so small that the flow rate has a much smaller effect on the amount of time spent exposed to the energy source.  Both AMS and PIT-MS measurements show that absolute abundances vary as a function of flow rate, but the relative abundances of products and general chemical trends are preserved across the range of flow rates investigated. For this reason, the data presented here were acquired using the optimal flow rate for each instrument but any differences in flow rate between instruments does not affect the conclusions.

\section{Results and Discussion}

\subsection{Gas Phase Composition (PIT-MS)}

\begin{figure}
\centering
\resizebox{5.5in}{!}
{\includegraphics{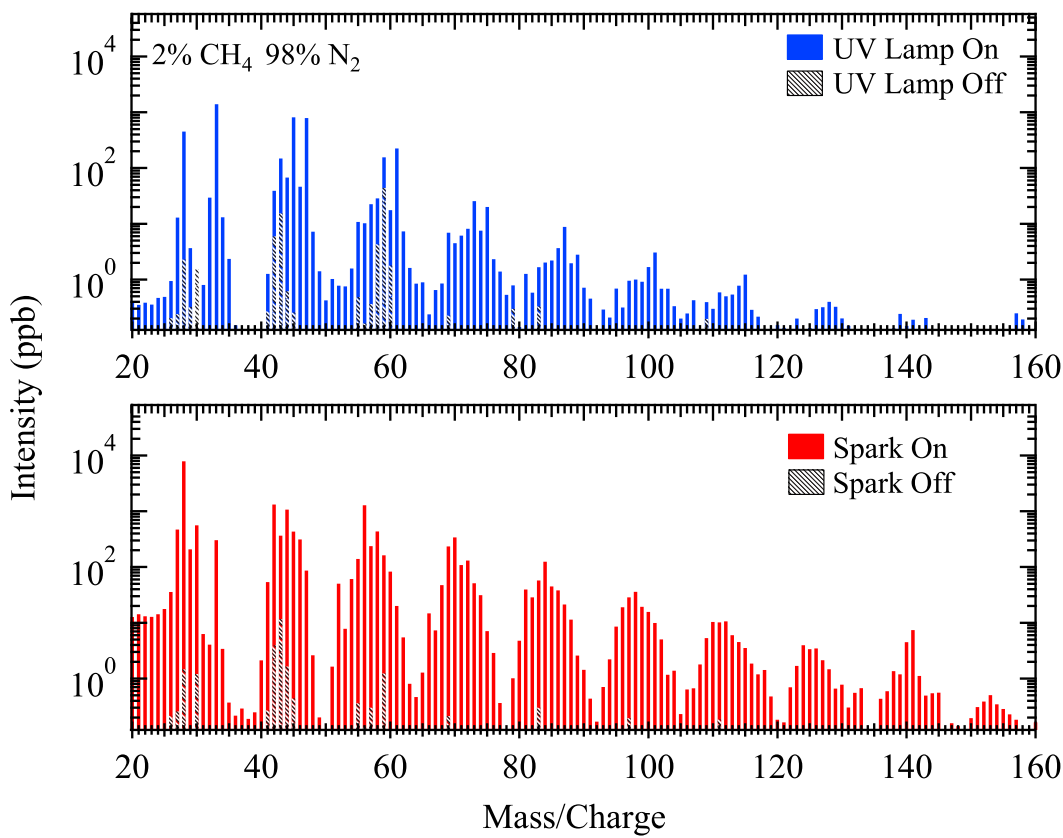}}
\caption{PIT-MS spectra of the gas phase products produced from 2\% CH$_{4}$ using either UV (blue, top) or spark (red, bottom) energy sources. Also shown are the spectra obtained when the energy source was off (shaded). The groups of peaks have a spacing of $\sim$14 mass units. The data presented here were obtained using the medium flow rate ($\sim$130 sccm). \label{fig:ptrms_on_off_2}}
\end{figure}

\begin{figure}
\centering
\resizebox{5.5in}{!}
{\includegraphics{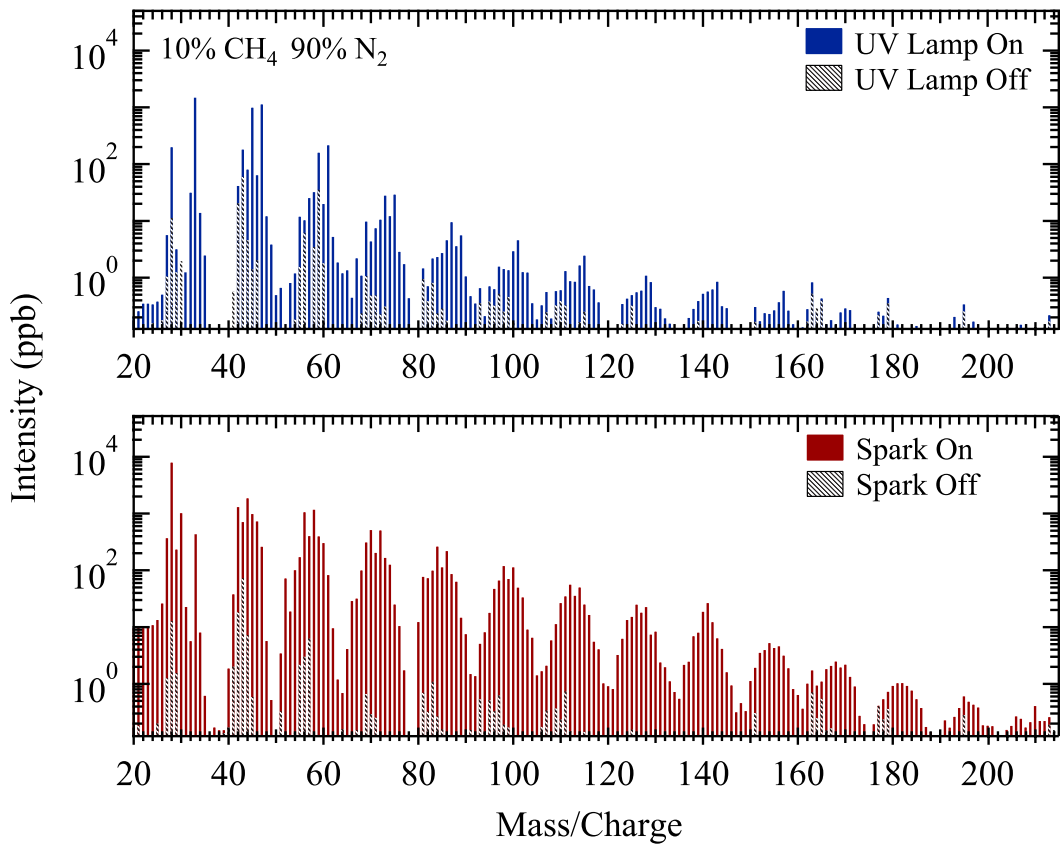}}
\caption{PIT-MS spectra of the gas phase products produced from 10\% CH$_{4}$ using either UV (blue, top) or spark (red, bottom) energy sources. Also shown are the spectra obtained when the energy source was off (shaded). The spark energy source resulted in the formation of gas phase species up to the mass limit of the instrument. To our knowledge, these are the heaviest gas phase products measured from a Titan atmosphere simulation experiment. The data presented here were obtained using the medium flow rate ($\sim$130 sccm). \label{fig:ptrms_on_off_10}}
\end{figure}

As shown in Figure \ref{fig:ptrms_on_off_2}, the mass spectra of the gas phase products are quite complex. For both the spark and UV experiments, products with masses greater than \emph{m/z} 100 were detected for every CH$_{4}$ concentration investigated; the 10\% CH$_{4}$ spark experiment produced products up to the mass limit of the instrument (217 amu/q) as shown in Figure \ref{fig:ptrms_on_off_10}. To our knowledge, these are the heaviest gas phase products ever detected from a Titan atmosphere simulation experiment. For a given initial CH$_{4}$ concentration, the spark experiments produced heavier gas phase products than the UV experiments. For both energy sources, the spectra exhibit groups of regularly spaced peaks with an average group spacing of $\sim$14, which is indicative of a difference of CH$_{2}$. Although the spectra shown here are averages of spectra taken over approximately 1 hr, the gas phase products exhibit very little time variation. Indeed the gas phase abundances stabilize within the first few minutes after turning on the energy source; unfortunately this prevents us from examining their behavior as a function of time. Note that there are still peaks present in the PIT-MS data (Figures \ref{fig:ptrms_on_off_2} and \ref{fig:ptrms_on_off_10}) when the energy source was off; these peaks represent no more than 1\% of the total signal. Inspection of the full set of experiments indicates they likely originate from the tubing connecting the chamber to the PIT-MS, despite the fact that it was pumped on over night between experiments.

\begin{figure}
\centering
\resizebox{5.7in}{!}
{\includegraphics{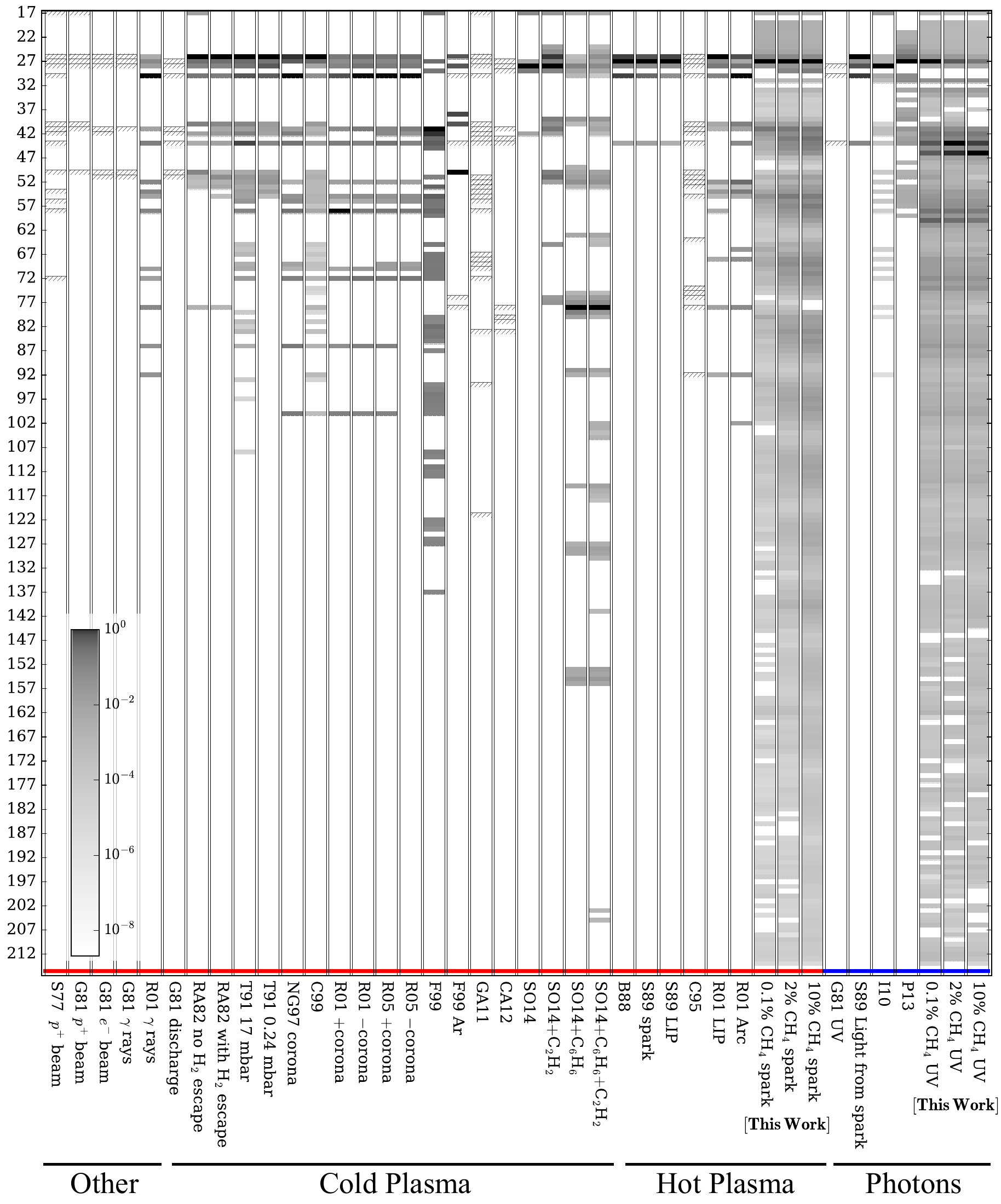}}
\caption{Comparison of previous measurements of gas phase products from Titan simulation experiments with the current work. For each work the reported values have been normalized to the most abundant product. All products with the same nominal mass have been summed. Hatched boxes indicate products that were reported without quantitative values. Additional information regarding these works can be found in Table \ref{table:mes}. \label{fig:all_gases}}
\end{figure}

\setlength{\tabcolsep}{4pt}
\begin{table}
\begin{center}
\caption{Summary of Titan Atmosphere Simulation Experiment Gas Phase Measurements\label{table:mes}}
\begin{tabular}{lllll}
\hline
Reference&&[CH$_{4}$]&Pressure&Energy Source \\
\hline
\citet{Scattergood:1977}&[S77]&50\% &1 bar&3.5 MeV protons\\
\citet{Gupta:1981}& [G81]&1\%&0.5 bar&UV (argon lamp)\\
&&1.3\%&0.75 bar&Discharge\\
&&1.4\% &0.71 bar&Gamma rays\\
&&3.8\% &0.002 bar&Proton beam\\
&&3.8\% &0.7 bar&Electron beam\\
\citet{Ramirez:2001}&[R01]&10\% &0.68 bar&Gamma rays\\
&&10\% &0.68 bar&LIP\\
&&10\% &0.68 bar&+ \& - Corona\\
&&10\% &0.68 bar&Arc\\
\citet{Raulin:1982}& [RA82]&33\% &0.2 bar&Discharge\\
\citet{Thompson:1991}& [T91]&10\% &0.02 bar&Cold plasma\\
&&10\% &0.02 bar&Cold plasma\\
Navarro-Gonz\'alez\nocite{Navarro:1997}& [NG97]&10\% &0.68 bar&Corona\\
\hspace{5 mm}and Ram\'irez (1997)&&&&\\

\citet{Coll:1999}& [C99]&2\%&0.002 bar&DC glow\\
\citet{Ramirez:2005}& [R05]&10\% &0.65 bar&+ \& - Corona\\
\citet{Fujii:1999}& [F99]&10\%$^{a}$&0.03 bar&Microwave discharge\\
\citet{Gautier:2011}& [GA11]&4\% &0.002 bar&RF discharge\\
\citet{Carrasco:2012}& [CA12]&1\% &10$^{-3}$ bar&RF discharge\\
\citet{Sciamma:2014}&[SO14]&10\%$^{b}$&0.03 bar&Cold plasma\\
\citet{Borucki:1988}& [B88]&3\% &1 bar&LIP\\
\citet{Scattergood:1989}& [S89]&10\%&1 bar&LIP\\
&&10\% &1 bar&Spark\\
&&10\% &1 bar&Photons from spark\\
\citet{Coll:1995}& [C95] &1.6\% &0.81 bar&Spark discharge\\
\citet{Imanaka:2010}&[I10]&5\% &10$^{-4}$ bar&60 nm photons\\
\citet{Peng:2013}&[P13]&$\sim$7.5\%$^{c}$&0.0087 bar&60-350 nm photons\\
\hline
\multicolumn{5}{l}{Unless otherwise noted, the bulk of the gas mixtures was N$_{2}$.}\\
\multicolumn{5}{l}{S77, SO14, I10, and P13 used mass spectrometry to measure products. F99 used}\\
\multicolumn{5}{l}{Li$^{+}$ attachment mass spectrometry. All other works listed here used GCMS}\\
\multicolumn{5}{l}{ (Gas Chromatography Mass Spectrometry). GA11, CA11, C99, and C95}\\
\multicolumn{5}{l}{collected products in a cold trap prior to analysis. $^{a}$F99 also ran an experiment}\\
\multicolumn{5}{l}{with Ar in place of N$_{2}$. $^{b}$SO14 also ran experiments with 10\% CH$_{4}$ 5\% C$_{2}$H$_{2}$,}\\
\multicolumn{5}{l}{10\% CH$_{4}$ 5\% C$_{6}$H$_{6}$, and 10\% CH$_{4}$ 5\% C$_{2}$H$_{2}$ 5\% C$_{6}$H$_{6}$.}\\
\multicolumn{5}{l}{$^{c}$P13 used 25\% He and ran additional flow rates, pressures, and CH$_{4}$ ratios.} \\
\end{tabular}
\end{center}
\end{table}

Figure \ref{fig:all_gases} compares the results of our experiment to a number of previous measurements of gas phase products of Titan atmosphere simulation experiments. Note that the experiments used a variety of gas mixtures and energy sources as well as a variety of different analytical methods. Due to the difference in analytical method used and differences in the type of information reported it is very difficult to directly compare the results of these experiments. We have normalized the results of each experiment to the most abundant product. We have also summed all products (including isomers when reported) for a given nominal mass. Some experiments were included although they did not report quantitative values (indicated by hatched boxes). While Figure \ref{fig:all_gases} does not contain every analysis ever performed, a number of general trends are evident. In general the abundance of products decreases with increasing mass. Additionally, HCN tends to be one of the most abundant products measured and has been observed in every experiment shown except for the experiment of \citet{Fujii:1999} when Ar was used instead of N$_{2}$. Aside from the work presented here, there exist very few measurements from UV experiments, making it difficult to compare the effect of energy source on the gas phase products. 

\begin{figure}
\centering
\resizebox{5.5in}{!}
{\includegraphics{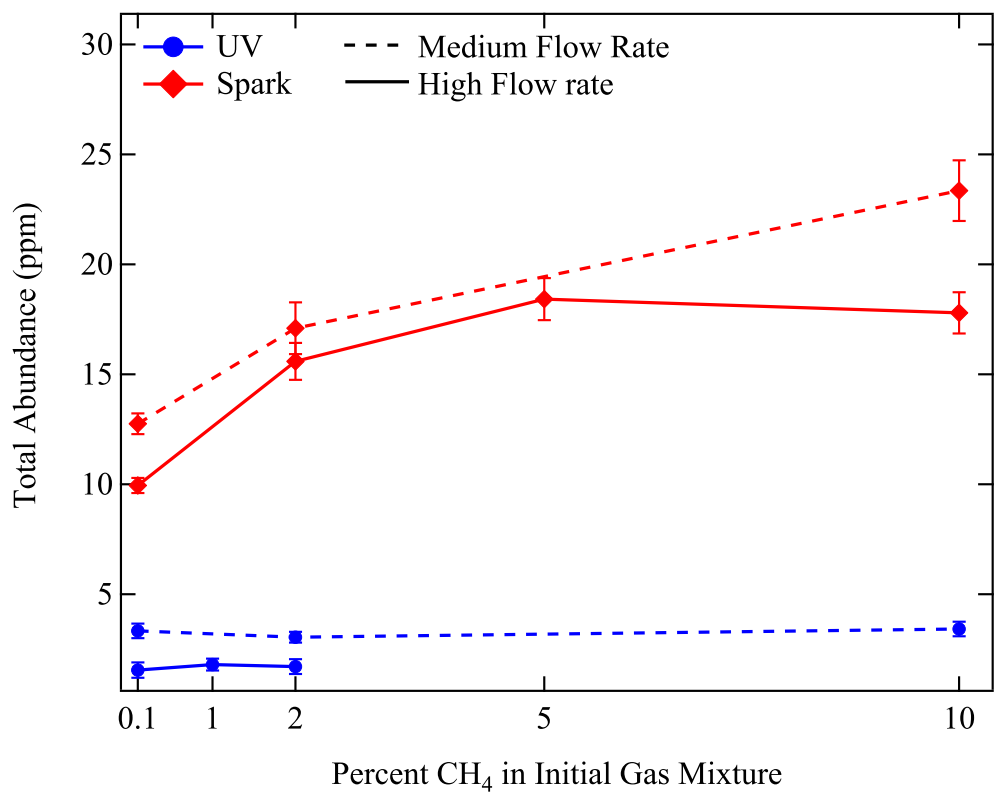}}
\caption{For both flow rates and all CH$_{4}$ mixing ratios, the spark experiments produce a greater total abundance of gas phase products than the UV experiments as shown in this plot of the total amount of gas phase products measured by the PIT-MS as a function of initial CH$_{4}$ concentration.\label{fig:ptrms_load}}
\end{figure}

Figure \ref{fig:ptrms_load} shows the sum of the intensity of all of the peaks in the spectra. The spark experiments always produce a higher total concentration of gas phase products than the UV experiments for a given CH$_{4}$ concentration. Additionally, the concentration of gas phase products generally increases with increasing CH$_{4}$ concentration for spark experiments. This trend was also observed by \citet{Gautier:2011}, and \citet{Carrasco:2012} in the measurements they obtained of the gas phase products from their cold plasma experiments. Unlike the spark experiments, the concentration of gas phase products from the UV experiments is relatively constant with initial CH$_{4}$ concentration. However, it is important to note that we cannot detect a number of small hydrocarbon species that we expect to be present in the experiment (e.~g. C$_{2}$H$_{2}$, C$_{2}$H$_{6}$, C$_{3}$H$_{8}$). The ability to detect these species would certainly increase the total gas phase abundances, but might also change the observed trends. For example, \citet{Carrasco:2012} observed a transition from nitrogen bearing species dominating the mass spectra below 5\% CH$_{4}$ to hydrocarbon species dominating above 5\% CH$_{4}$.

\subsubsection{Comparison to Titan's atmospheric composition}

\setlength{\tabcolsep}{6pt}
\renewcommand{\arraystretch}{0.70}
\begin{longtable}{lllll}
\caption{Composition of Titan's Atmosphere Compared to PIT-MS Measurements of 2\% CH$_{4}$ Spark and UV Experiments\label{table:comp}}\\
\hline
Formula&UV&Spark&Stratosphere$^{(a)}$&Thermosphere$^{(b,c)}$\\ \hline
\endfirsthead
\hline
\endhead
\hline \multicolumn{3}{r}{\textit{Continued on next page}}
\endfoot
\hline
\endlastfoot
\multicolumn{5}{l}{Molecules that can be detected (proton affinity higher than H$_{2}$O)}\\
\hline
\textbf{CH$_{3}$C$_{2}$H}&1$\times10^{-9}$&5$\times10^{-8}$&4.8$\times10^{-9}$&1.4$\times10^{-4}$\\
C$_{3}$H$_{6}$&1$\times10^{-7}$&4$\times10^{-7}$&2.5$\times10^{-9}$$^{(d)}$&2.3$\times10^{-6}$$^{(e)}$\\ 
\textbf{C$_{4}$H$_{2}$}&1$\times10^{-9}$&2$\times10^{-9}$&1.12$$$\times10^{-9}$&6.4$\times10^{-5}$-1.0$\times10^{-5}$\\
\textbf{C$_{6}$H$_{6}$}&8$\times10^{-10}$&1$\times10^{-9}$&2.2$\times10^{-10}$ &8.95$\times10^{-7}$-3.0$\times10^{-6}$\\
\textbf{HCN}&5$\times10^{-7}$&8$\times10^{-6}$&6.7$\times10^{-8}$&2.0$\times10^{-4}$\\
\textbf{HC$_{3}$N}&8$\times10^{-10}$&5$\times10^{-8}$&2.8$\times10^{-10}$&3.2$\times10^{-5}$-4.0$\times10^{-5}$\\
HC$_{5}$N&2$\times10^{-9}$&3$\times10^{-9}$&&	1.0$\times10^{-6}$\\
\textbf{CH$_{3}$CN}&4$\times10^{-8}$&1$\times10^{-6}$&8$\times10^{-9}$$^{(f)}$&3.1$\times10^{-5}$-3.0$\times10^{-6}$\\
\textbf{C$_{2}$H$_{3}$CN}&2$\times10^{-9}$&6$\times10^{-8}$&$<$2$\times10^{-9}$$^{(f)}$&$<$1.8$\times10^{-5}$-1.0$\times10^{-5}$\\
C$_{2}$H$_{5}$CN&1$\times10^{-8}$&1$\times10^{-6}$&$1.3\times10^{-8}$$^{(g)}$&5.0$\times10^{-7}$\\
C$_{5}$H$_{5}$N&1$\times10^{-10}$&5$\times10^{-9}$&&4.0$\times10^{-7}$\\
\textbf{C$_{6}$H$_{7}$N}&2$\times10^{-10}$&2$\times10^{-9}$&&1.0$\times10^{-7}$\\
\textbf{CH$_{2}$NH}&ND&6$\times10^{-7}$&&1.0$\times10^{-5}$\\
\textbf{NH$_{3}$}	&*&*&$<$1.9$\times10^{-10}$$^{(h)}$& 6.7$\times10^{-6}$\\
\hline
\multicolumn{5}{l}{Molecules with unknown proton affinity}\\
\hline
C$_{6}$H$_{2}$&2$\times10^{-8}$&7$\times10^{-9}$&$<$6.0$\times10^{-10}$$^{(i)}$&8.0$\times10^{-7}$\\ 
C$_{7}$H$_{4}$&3$\times10^{-9}$&3$\times10^{-9}$&&3.0$\times10^{-7}$\\ 
C$_{8}$H$_{2}$&9$\times10^{-10}$&2$\times10^{-8}$&&	2.0$\times10^{-7}$\\ 
\textbf{C$_{4}$H$_{3}$N}&2$\times10^{-10}$&2$\times10^{-8}$&&4.0$\times10^{-6}$\\ 
C$_{6}$H$_{3}$N&7$\times10^{-10}$&1$\times10^{-9}$&&3.0$\times10^{-7}$\\ 
\hline
\multicolumn{5}{l}{Molecules that cannot be detected (proton affinity lower than H$_{2}$O)}\\
\hline
CH$_{4}$&&&1.6$\pm0.5\%$$^{(j)}$&2.20\%\\
H$_{2}$&&&9.6$\pm2.4$$\times10^{-4}$$^{(k)}$&3.9$\times10^{-3}$\\
C$_{2}$H$_{2}$&&&2.97$\times10^{-6}$&3.1$\times10^{-4}$\\
C$_{2}$H$_{4}$&&&1.2$\times10^{-7}$&3.1$\times10^{-4}$-1.0$\times10^{-3}$\\
C$_{2}$H$_{6}$&&&7.3$\times10^{-6}$&7.3$\times10^{-5}$\\
C$_{3}$H$_{8}$&&&4.5$\times10^{-7}$ &$<$4.8$\times10^{-5}$\\
C$_{2}$N$_{2}$&&&$<$1.0$\times10^{-9}$$^{(l)}$&4.8$\times10^{-5}$\\ 
\hline
\multicolumn{5}{l}{All values come from the work referenced in the header of the table}\\
\multicolumn{5}{l}{unless otherwise noted in the table. }\\
\multicolumn{5}{l}{\textbf{Bold} indicates that the molecule is the only possibility at that nominal}\\
\multicolumn{5}{l}{mass in the PIT-MS measurements}\\
\multicolumn{5}{l}{*Our use of H$_{3}$O$^{+}$ as the primary ion results in difficulties with}\\
\multicolumn{5}{l}{quantification of NH$_{3}$ }\\
\multicolumn{5}{l}{$^{(a)}$\citet{Coustenis:2010} (at 5$^{\circ}$S; HCN, C$_{6}$H$_{6}$, C$_{3}$H$_{8}$, C$_{4}$H$_{2}$, and HC$_{3}$N }\\
\multicolumn{5}{l}{exhibit latitudinal variations \citep{Coustenis:2007}) values averaged} \\
\multicolumn{5}{l}{over TB-T44 assuming constant vertical profiles}\\
\multicolumn{5}{l}{$^{(b)}$\citet{Cui:2009} (at 1077 km INMS Closed Source Neutral mode,}\\
\multicolumn{5}{l}{signals from C$_{2}$H$_{2}$ and C$_{2}$H$_{4}$ are difficult to separate so the reported}\\
\multicolumn{5}{l}{value is for both species combined)}\\
\multicolumn{5}{l}{$^{(c)}$\citet{Vuitton:2007} (1100 km INMS Open Source mode)}\\
\multicolumn{5}{l}{$^{(d)}$\citet{Nixon:2013b} (at 225 km)}\\
\multicolumn{5}{l}{$^{(e)}$\citet{Magee:2009} (global average at 1050 km)}\\
\multicolumn{5}{l}{$^{(f)}$\citet{Marten:2002} (at 200 km) $^{(g)}$\citet{Cordiner:2015} (at 300 km)}\\
\multicolumn{5}{l}{$^{(h)}$\citet{Delpech:1994}}\\
\multicolumn{5}{l}{$^{(i)}$\citet{Teanby:2013} (3-$\sigma$ upper limit, peak sensitivity at 75 km)}\\
\multicolumn{5}{l}{$^{(j)}$\citet{Flasar:2005} $^{(k)}$\citet{Courtin:2007} $^{(l)}$\citet{Delpech:1994}}\\
\end{longtable}
\renewcommand{\arraystretch}{1}

\begin{figure}
\centering
\resizebox{5.5in}{!}
{\includegraphics{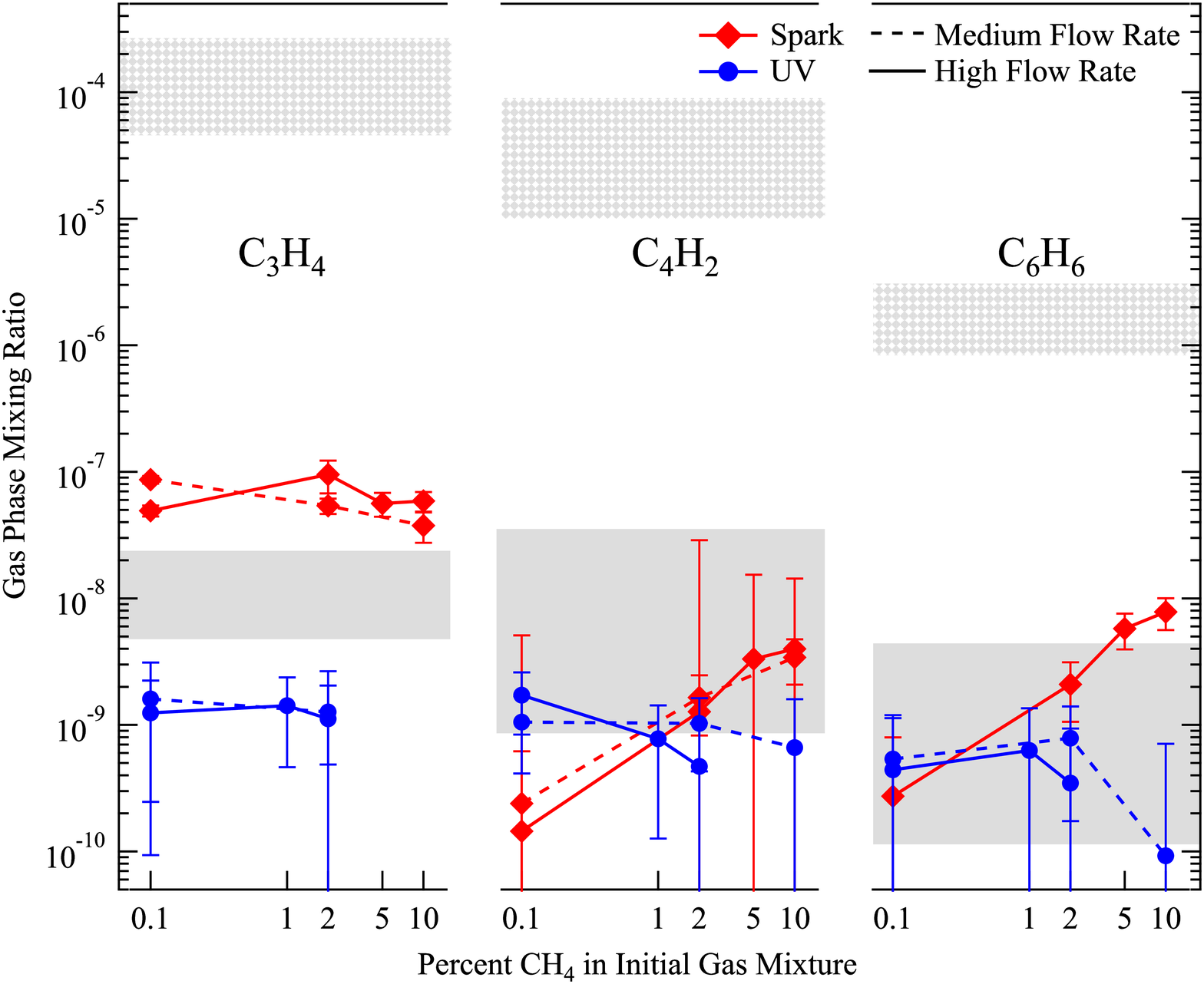}}
\caption{Comparison of the C$_{3}$H$_{4}$, C$_{4}$H$_{2}$, and C$_{6}$H$_{6}$ abundances in the spark (red diamonds) and UV (blue circles) experiments as a function of initial CH$_{4}$ abundance to the abundances in Titan's atmosphere. Dashed lines indicate medium flow rate experiments and solid lines indicate high flow rate experiments. The solid gray areas indicate measured abundances in Titan's stratosphere from   Cassini CIRS \citep{Coustenis:2010} (range indicates latitude variation). The patterned areas indicate measured abundances in Titan's thermosphere from Cassini INMS (range includes values from \citet{Vuitton:2007} and \citet{Cui:2009} at 1100 km and 1077 km).
\label{fig:cgases}}
\end{figure}

\begin{figure}
\centering
\resizebox{5.5in}{!}
{\includegraphics{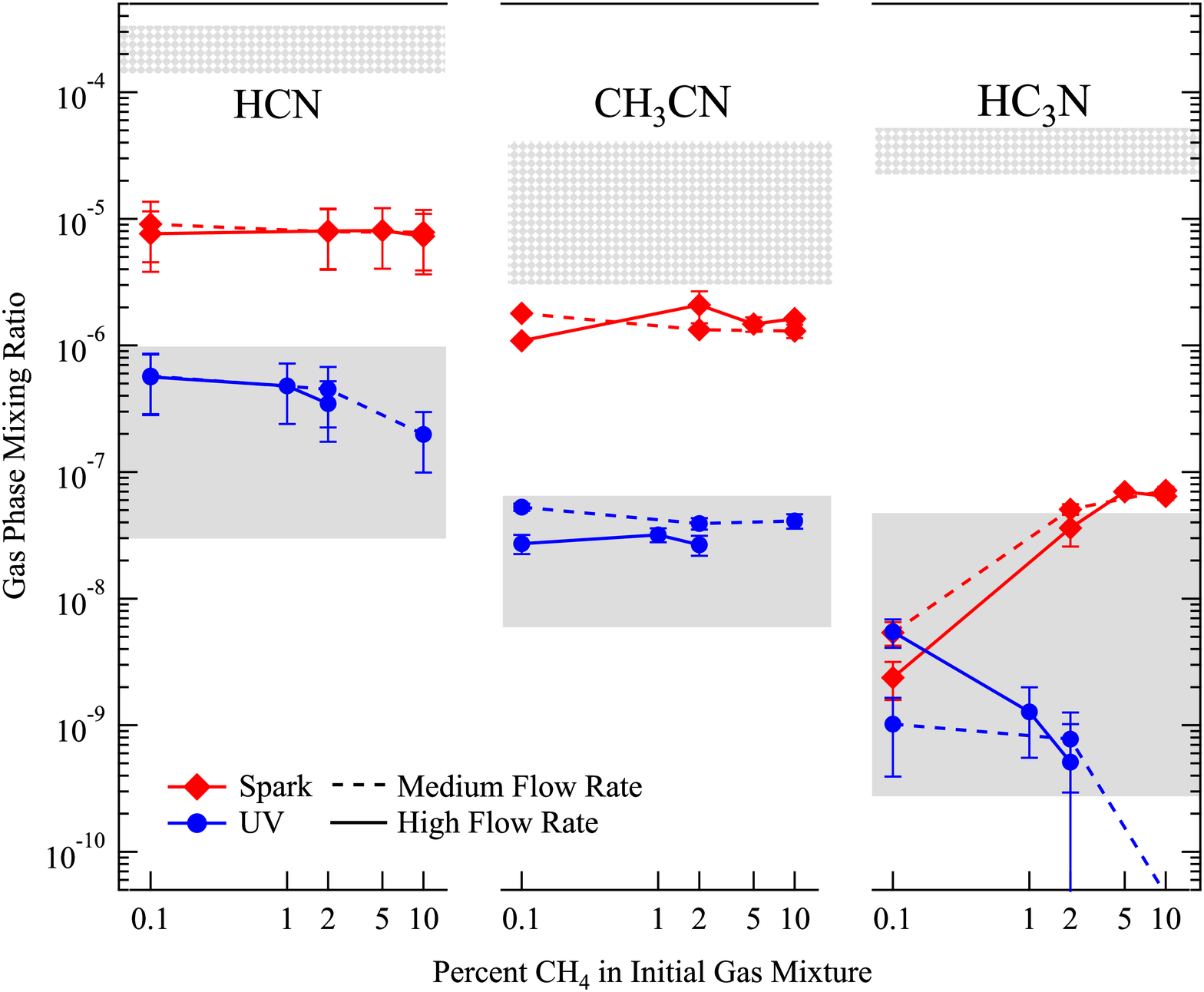}}
\caption{Comparison of the HCN, CH$_{3}$CN, and HC$_{3}$N abundances in the spark (red diamonds) and UV (blue circles) experiments as a function of initial CH$_{4}$ abundance to the abundances in Titan's atmosphere. Dashed lines indicate medium flow rate experiments and solid lines indicate high flow rate experiments. The solid gray areas indicate measured abundances in Titan's stratosphere; the HCN and HC$_{3}$N measurements are from Cassini CIRS \citep{Coustenis:2010} (range indicates latitude variation) and the CH$_{3}$CN abundance is from ground-based measurements \citep{Marten:2002} (range indicates altitude variation). The patterned areas indicate measured abundances in Titan's thermosphere from Cassini INMS (range includes values from \citet{Vuitton:2007}, \citet{Cui:2009}, and \citet{Magee:2009} at 1100 km, 1077 km, and 1050 km respectively).
\label{fig:ngases}}
\end{figure}

Due to the unit mass resolution of the instrument, most peaks cannot be uniquely identified based on mass alone. However, the requirement for the molecule to have a higher proton affinity than water eliminates a number of possibilities at the lower masses allowing for some unique identifications. Table \ref{table:comp} compares the abundances measured for the 2\% CH$_{4}$ spark and UV experiments to the composition of Titan's stratosphere and thermosphere. It also indicates which molecules present in Titan's atmosphere cannot be detected using this technique as discussed in Section \ref{sect:pitms}. Figures \ref{fig:cgases} and \ref{fig:ngases} show the abundances of a 3 hydrocarbons (C$_{3}$H$_{4}$, C$_{4}$H$_{2}$, and C$_{6}$H$_{6}$) and 3 nitriles (HCN, CH$_{3}$CN, and HC$_{3}$N), respectively, observed in our spark and UV experiments as a function of initial CH$_{4}$ concentration compared to the abundances of those three molecules observed in Titan's stratosphere and thermosphere from the Cassini Composite Infrared Spectrometer (CIRS) \citep{Coustenis:2010}, ground-based observations \citep{Marten:2002}, and the Cassini Ion and Neutral Mass Spectrometer (INMS) \citep{Vuitton:2007, Cui:2009, Magee:2009}. The six molecules shown in Figures \ref{fig:cgases} and \ref{fig:ngases} are identified based on mass and proton affinity, however only acetonitrile (CH$_{3}$CN) and benzene (C$_{6}$H$_{6}$) have been specifically calibrated for the PIT-MS \citep{Warneke:2011}. Other species exist with those masses but their proton affinities are not high enough to be plausible candidates. Remarkably, the abundances measured in our experiments are nearly identical to the abundances observed in Titan's atmosphere for the UV experiments, with the exception of C$_{3}$H$_{4}$. Ethyl cyanide (C$_{2}$H$_{5}$CN) has recently been definitively detected in Titan's stratosphere for the first time \citep{Cordiner:2015} and our UV abundance is consistent with that measurement as well. Both the wavelengths of the FUV photons available to drive the chemistry in this experiment and the pressure used in these experiments are roughly analogous to Titan's stratosphere (see e.g., \citet{Horst:2017}). Although some of these species are produced in the ionosphere, rather than the stratosphere, these results indicate that our UV experiments produce particles in an environment that is roughly similar to Titan (with the notable exception of temperature). While our post-Cassini understanding of Titan haze formation indicates that particle formation begins in the ionosphere, particle nucleation, growth, and evolution may also occur in the stratosphere driven by longer wavelength photons. The abundances in the spark experiments are uniformly too low, particularly for the hydrocarbon species, indicating that we are producing nitrogen bearing species relatively more efficiently than in Titan's atmosphere. However, the relative abundances of the smaller species follow the same general trend as observed in Titan's atmosphere (less acetonitrile than hydrogen cyanide, for example). Given that the pressure in our experiments is higher than the pressure in Titan's thermosphere it is perhaps not surprising that the abundances are not correct.   

In addition to the molecules shown in Figures \ref{fig:cgases} and \ref{fig:ngases}, we also observe methanimine (CH$_{2}$NH) in the spark experiments, but not in the UV experiments. CH$_{2}$NH has been detected in Titan's thermosphere \citep{Vuitton:2007, Yelle:2010}, but not the stratosphere. \citet{Carrasco:2012} detected CH$_{2}$NH in the gas phase products of their radio frequency cold plasma experiments and suggested that the need for N($^{2}$D) for production of CH$_{2}$NH means that CH$_{2}$NH is a ``characteristic product'' of plasma experiments. In Titan's atmosphere, CH$_{2}$NH is produced mainly by two reactions NH+CH$_{3} \rightarrow$ CH$_{2}$NH+H and N($^{2}$D)+CH$_{4} \rightarrow$ CH$_{2}$NH+H \citep{Yelle:2010}. The absence of CH$_{2}$NH in our UV experiments is not surprising given the lack of N($^{2}$D), but the apparent absence of NH may provide insight into the mechanism responsible for N$_{2}$ dissociation.

Of the main peaks identified in the plasma experiments of \citet{Carrasco:2012} (HCN, CH$_{3}$CN, C$_{2}$H$_{3}$CN, C$_{2}$H$_{5}$CN, HC$_{3}$N and C$_{2}$N$_{2}$), we detect all but C$_{2}$N$_{2}$, which has a proton affinity lower than H$_{2}$O and therefore cannot be detected using this technique. A cold trap was employed on the same experiment by \citet{Gautier:2011} and they detected almost 40 gas phase products using GC-MS. With the exception of the small hydrocarbon species that we cannot detect, we have signal at the nominal masses of each of the molecules they observe. However, due to the resolution of the PIT-MS and the inability of mass spectrometry to distinguish between isomers, we cannot directly compare our gas phase measurements to theirs. We also detect all of the ions in the mass spectrum obtained by \citet{Imanaka:2010} from VUV/EUV Titan atmosphere simulation experiments that we are capable of detecting, although the mixing ratios produced by both our spark and UV experiments are much lower than theirs. Surprisingly, the relative ratios of HCN, CH$_{3}$CN, and C$_{7}$H$_{8}$ to C$_{6}$H$_{6}$ (molecules we can definitively identify) in our UV experiments are nearly identical to the ratios observed by \citet{Imanaka:2010} for experiments performed using 60 nm photons!

\subsubsection{Gas phase nitrogen incorporation}

For the purposes of this work, we are particularly interested in the degree of nitrogen incorporation in the gas phase species. However since we cannot uniquely identify all of the peaks in the data, it is not possible to calculate the absolute abundance of nitrogen in all of the gas phase products. Additionally, as previously discussed we are almost certainly missing a large fraction of the carbon bearing species. All of the nitrogen bearing species that have been detected in Titan's atmosphere, with the exception of N$_{2}$, have proton affinities higher than H$_{2}$O and can be detected in our measurements. Taken together, the biases and constraints of this technique result in the placement of an upper limit on the amount of nitrogen in the gas phase based on these measurements.

\begin{figure}
\centering
\resizebox{5.5in}{!}
{\includegraphics{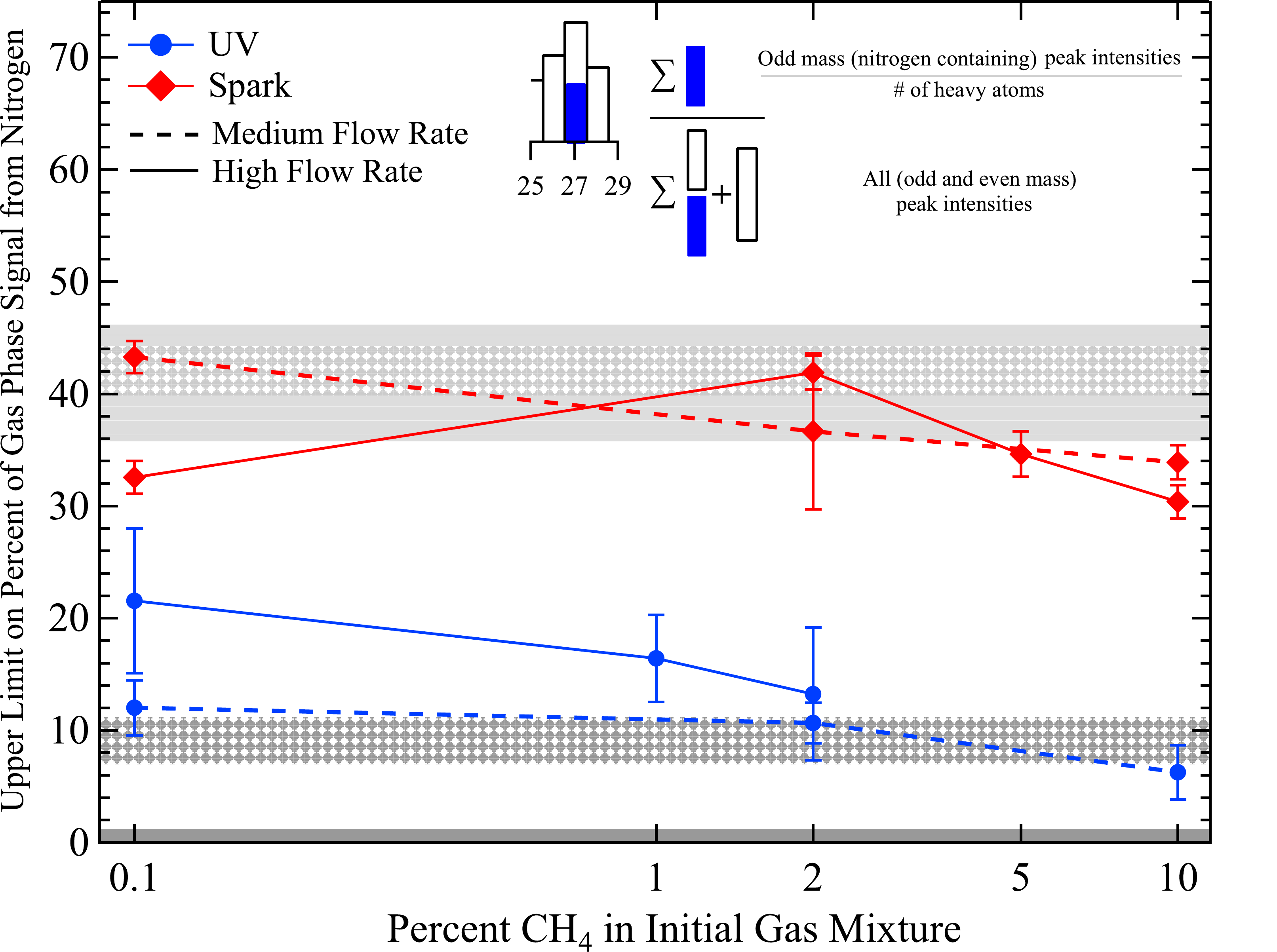}}
\caption{Shown here is an upper limit on the total mixing ratio of nitrogen in the gas phase products of the experiment. We cannot calculate the absolute percent of gas phase signal coming from nitrogen as these PIT-MS measurements are biased against small hydrocarbon bearing species, which we expect to be produced in large abundances in our experiment. The dark gray shaded regions indicate the values one would obtain doing the exact same calculation for Titan's stratosphere (solid) and thermosphere (patterned) using Cassini CIRS \citep{Coustenis:2010, Nixon:2013b}, ground-based \citep{Marten:2002}, ALMA \citep{Cordiner:2015}, and Cassini INMS \citep{Vuitton:2007, Cui:2009, Magee:2009} measurements, except for CH$_{4}$, N$_{2}$, noble gases, and oxygen bearing species. The light gray shaded regions is the same calculation but with any molecules that could not be detected by our PIT-MS measurements removed. \label{fig:ptrms_n}}
\end{figure}

To estimate an upper limit on the degree of nitrogen incorporation in the gas phase, we take advantage of the nitrogen rule, which states that a molecule that is composed of an odd number of N and any number of C, H, O, S, Si, P, or halogens will have an odd nominal mass. This rule results from the unique fact that nitrogen has an odd valence (+3) but an even nominal mass (14) (see e.g. \citet{Watson:2007}). For example, HCN has an odd nominal mass (27) and 1 nitrogen atom. Accordingly, molecules with an even nominal mass have either no nitrogen or an even number of nitrogen atoms. For this analysis, we assume that molecules with odd nominal masses (corresponding to even masses in the PIT-MS data due to the addition of the proton) have one nitrogen atom and molecules with an even nominal mass contain no nitrogen. This is not a perfect assumption since molecules may contain 2 or more nitrogen atoms; however, because abundance generally decreases with increasing mass, the lower mass peaks, which most likely contain zero or one nitrogen atom, contribute the most signal. We have not accounted the $^{13}$C contribution to odd peaks as it does not significantly alter the results.  Figure \ref{fig:ptrms_n} shows the upper limits on the amount of gas phase nitrogen incorporation in the experiments. For all initial concentrations of CH$_{4}$, the upper limit on nitrogen content is higher for the spark than the UV, which is perhaps not surprising given that the FUV energy source cannot directly dissociate nitrogen while the electrical discharge can.  In both the spark and UV experiments, the nitrogen content generally decreases with increasing CH$_{4}$ concentration, mostly likely a consequence of the introduction of more carbon into the system. Figure \ref{fig:ptrms_n} also shows a comparison of the PIT-MS measurements to Titan's atmosphere, using the same data that were used in Figures \ref{fig:cgases} and \ref{fig:ngases}. Here two different comparisons are shown. The light colored solid and patterned regions indicate the values for Titan's atmosphere if only the molecules that could be detected using the PIT-MS were used in the calculation. There is remarkable agreement between the spark measurements and the value for Titan's thermosphere calculated in the same way. The UV experiments demonstrate much poorer agreement, which is potentially not surprising given the inability of FUV photons to directly dissociate N$_{2}$. The dark solid and patterned regions indicate the actual values for Titan's atmosphere including all molecules that have been detected except for N$_{2}$ and CH$_{4}$. The difference indicates the level of bias that is introduced by our measurement technique. 

\subsubsection{Gas phase functional groups}

\begin{figure}
\centering
\resizebox{5.5in}{!}
{\includegraphics{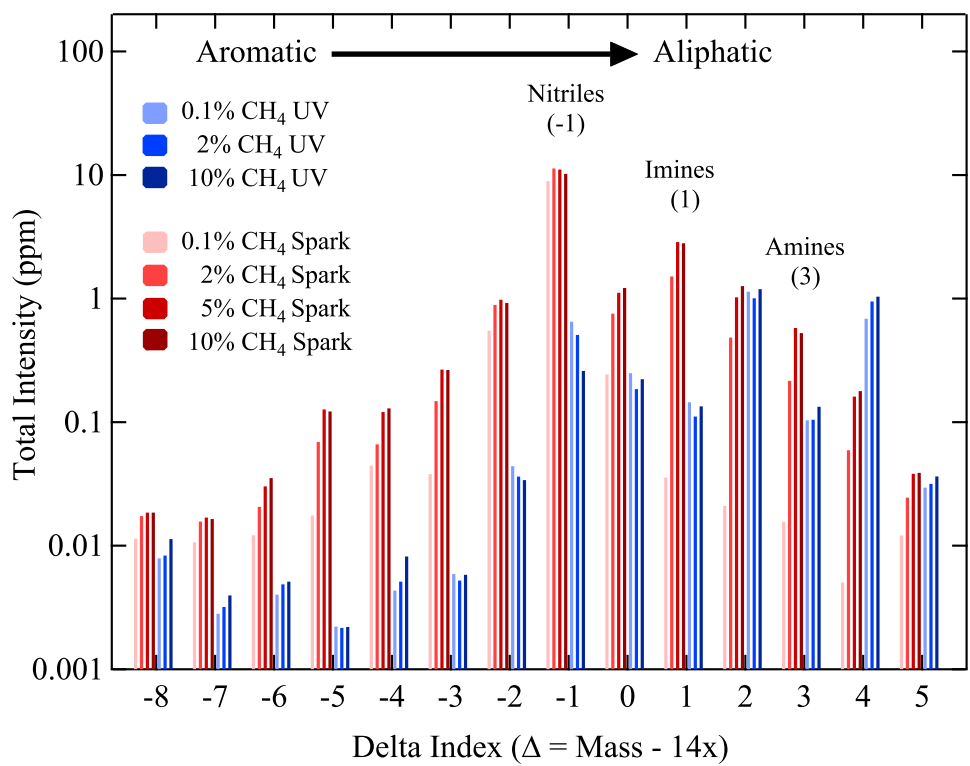}}
\caption{Abundance distribution as a function of delta index ($\Delta =$ Mass$- 14x$) for gas phase products from the spark (red) and UV (blue) experiments for a range of initial CH$_{4}$ concentrations. The data presented here were obtained using the medium flow rate ($\sim$130 sccm). Positive delta indices correspond to a greater degree of saturation in the molecules. Odd delta indices indicate the presence of an odd number of nitrogen; even delta indices correspond to the absence of nitrogen or an even number of nitrogen atoms.  \label{fig:delta}}
\end{figure}

Although we cannot uniquely identify all of the molecules due to the resolution of the PIT-MS, we can use the technique of delta analysis to probe the possible functional groups present and the degree of saturation of the molecules observed. A delta index for each peak is calculated by subtracting units of 14 (corresponding to CH$_{2}$) from the nominal mass. The standard formula is given by $\Delta =$ Mass $-14x+1$, however this formula must be modified for use with soft ionization techniques because we are looking at molecules, not fragments. Therefore, we define $\Delta =$ Mass $-14x$ (see e.g. \citet{Mclafferty:1993}). Using this definition, odd indices correspond to species containing an odd number of nitrogen atoms, while even indices correspond to zero or an even number of nitrogen. This technique is generally used for molecules with a single functional group connected to a saturated aliphatic substructure and so it becomes less diagnostic at higher masses where more than one nitrogen atom may be present. However, since the majority of the signal in the PIT-MS data corresponds to low mass species, this method is still instructive. Also, the regular spacing of 14 amu indicates that the assumption of saturated aliphatic substructure for the purposes of this calculation is not unreasonable. Shown in Figure \ref{fig:delta} is a comparison of the delta indices for the spark and UV experiments for a range of initial CH$_{4}$ concentrations. In general the spark experiments have higher total intensities at each delta index because they result in higher total gas phase species abundances. On average, the spark experiments have delta indices that correspond to a slightly lower degree of saturation (a higher degree of aromaticity) than the UV experiments. Perhaps one of the most striking results is the how the gas phase abundances of nitrogen bearing species differ as a function of saturation. Comparison of peaks with $\Delta$= -1, 1, 3 (corresponding to nitriles, imines, and amines respectively) shows that abundances of nitrogen bearing species decrease with increasing saturation for both the spark and UV experiments, although the trend is more pronounced for the spark experiments. This result is consistent with previous Titan atmosphere simulation experiments, which have shown that gas phase nitrogen bearing species are predominantly nitriles \citep{Thompson:1991, Coll:1995, Coll:1999, Gautier:2011}. This trend has also been observed in Titan's atmosphere where the majority of nitrogen bearing species detected are nitriles. While it is tempting to look at similar series of carbon bearing species such as alkynes, alkenes, and alkanes (-2,0,2 respectively) our inability to detect the low mass members of the groups, which are presumably the most abundant, might lead us astray. We must also be careful when interpreting this series because oxygen bearing molecules, resulting from contamination, are often found at these $\Delta$ values (e.g. $\Delta$=2 for aldehydes and ketones). 

\begin{figure}
\centering
\resizebox{5.5in}{!}
{\includegraphics{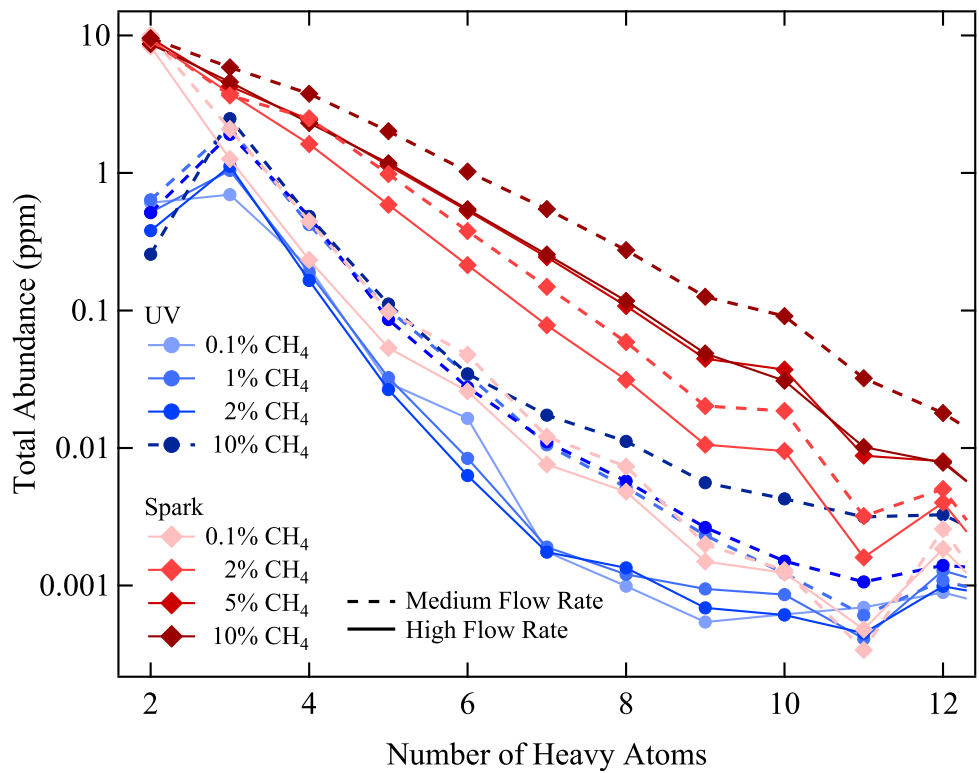}}
\caption{Total abundance measured (ppm) as a function of the total number of heavy atoms. Dashed lines indicate medium flow rate experiments and solid lines indicate high flow rate experiments. Error bars have not been included for the sake of clarity.\label{fig:heavy_gas}}
\end{figure}

\citet{Gautier:2011} looked specifically at the gas phase nitriles produced in their experiment and proposed a power law fit to their data where nitrile concentration is given by [C$_{x}$H$_{2x-1}$N$]=100x^{-5}$ based on 5 nitriles ranging from C=1 to C=4 (with C=4 using 2 isomers) \citet{Gautier:2011}. We can perform the same calculation and although we cannot differentiate between isomers, they summed the signal from their isomers which has the same effect. However, we observe a different power law with the exponent ranging from approximately -4 to -1.9 for the spark experiments and -3.8 to -2.8 for the UV experiments. Unlike \citet{Gautier:2011}, we observe that the power is a function of initial CH$_{4}$ concentration with the exponent increasing (becoming less negative) with increasing CH$_{4}$ concentration for both the spark and UV. It should be noted, however, that power law behavior as a function of the number of heavy atoms is observed for the total spectra as well (as shown in Figure \ref{fig:heavy_gas}). However, when all of the peaks are taken into account, the exponents are larger, ranging from -5.4 to -3.5 for the spark and -7.2 to -5.7 for the UV, again as a function of CH$_{4}$ concentration. Interestingly, while power law fits are extremely good for the UV data, the spark data are actually fit better using exponentials with exponents ranging from -0.9 to -0.6 as a function of gas mixture. Though it is tempting to perform this type of calculation for Titan's atmosphere, the various instrument biases, detection limits, and mass range limits make such a comparison difficult to interpret. Even for nitriles, the observational constraints are relatively poor. Only three nitriles have been definitively detected-- HCN (see e.g. \citet{Coustenis:2007, Coustenis:2010}), which has a very strong latitudinal variation, acetonitrile (CH$_{3}$CN) \citep{Marten:2002}, which was a disk averaged measurement, and ethyl cyanide (C$_{2}$H$_{5}$CN) \citep{Cordiner:2015}. For the three nitriles detected, the abundances are fit well by a power law with an exponent of $\sim$-2.7 (using values at 200 km from the previous references) which is consistent with the trend observed in our UV experiments. 

We cannot emphasize enough the unexplained presence of nitrogen bearing species in the gas phase products of the UV experiments. As mentioned in Section 2, previous works have demonstrated dissociation of nitrogen using an FUV energy source that is incapable of directly dissociating N$_{2}$ \citep{Dodonova:1966, Scattergood:1989, Hodyss:2011,Trainer:2012, Yoon:2014}. Here we definitively demonstrate the presence of nitrogen incorporation in the gas phase products. Since the mechanism responsible for nitrogen incorporation in these experiments is not yet understood, we do not yet know if the mechanism is important in Titan's atmosphere. However, it is possible that an important reaction or set of reactions is missing from current models of Titan's atmosphere. 



\subsection{Aerosol Phase Composition (HR-ToF-AMS)}

\begin{figure}
\centering
\resizebox{5.5in}{!}
{\includegraphics{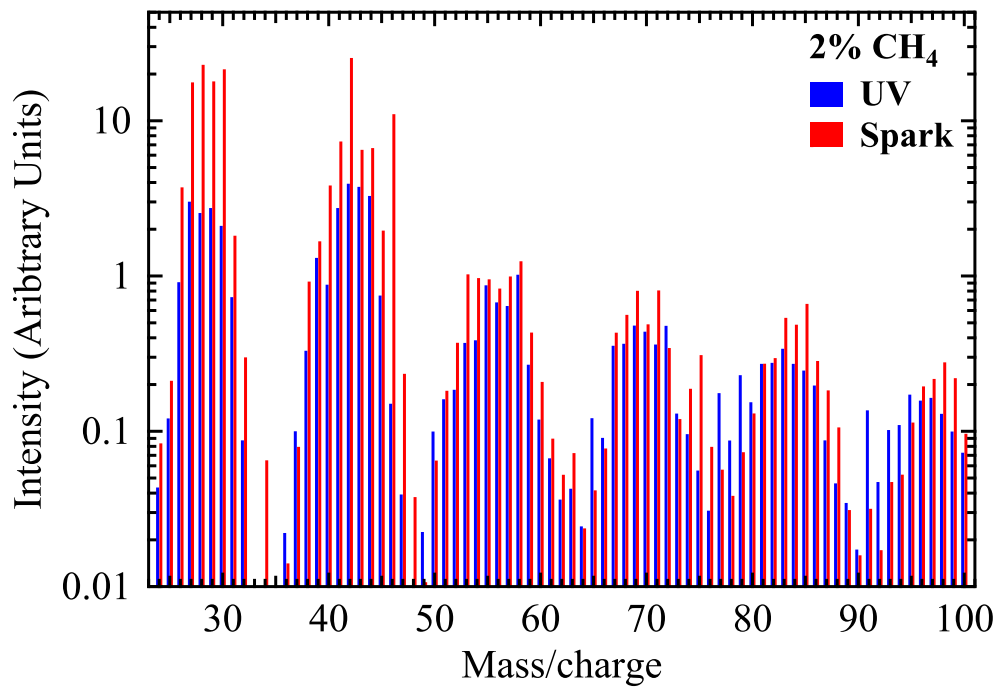}}
\caption{Comparison of AMS mass spectra for the 2\% initial CH$_{4}$ concentration spark (red) and UV (blue) experiments. Note that the setting used on the Tesla coil (spark experiments) was lower than previously published works using this experimental setup \citep{Trainer:2004b, Trainer:2004}, which resulted in a decrease in the amount of aromatics produced. \label{fig:ams2per}}
\end{figure}

A comparison of the AMS mass spectra for aerosol produced from 2\% CH$_{4}$ using spark and UV energy sources is shown Figure \ref{fig:ams2per}. In general, the shape of the groups of peaks and the spacing of the groups of peaks are quite similar for the two different energy sources. Notable differences include the apparent preference for smaller mass ions in the spark samples and the larger abundance of aromatic peaks (\emph{m/z} 77 and 91) in the UV samples. Previous experiments performed using a similar experimental setup at the University of Colorado demonstrated significant aromatic molecule production using the spark energy source \citep{Trainer:2004b, Trainer:2004}, contrary to what we observe in the current experiments. The previous experiments used the higher voltage setting of the Tesla coil and preliminary tests indicate that is the source of the differences between the current and previous works. 

\begin{figure}
\centering
\resizebox{5.5in}{!}
{\includegraphics{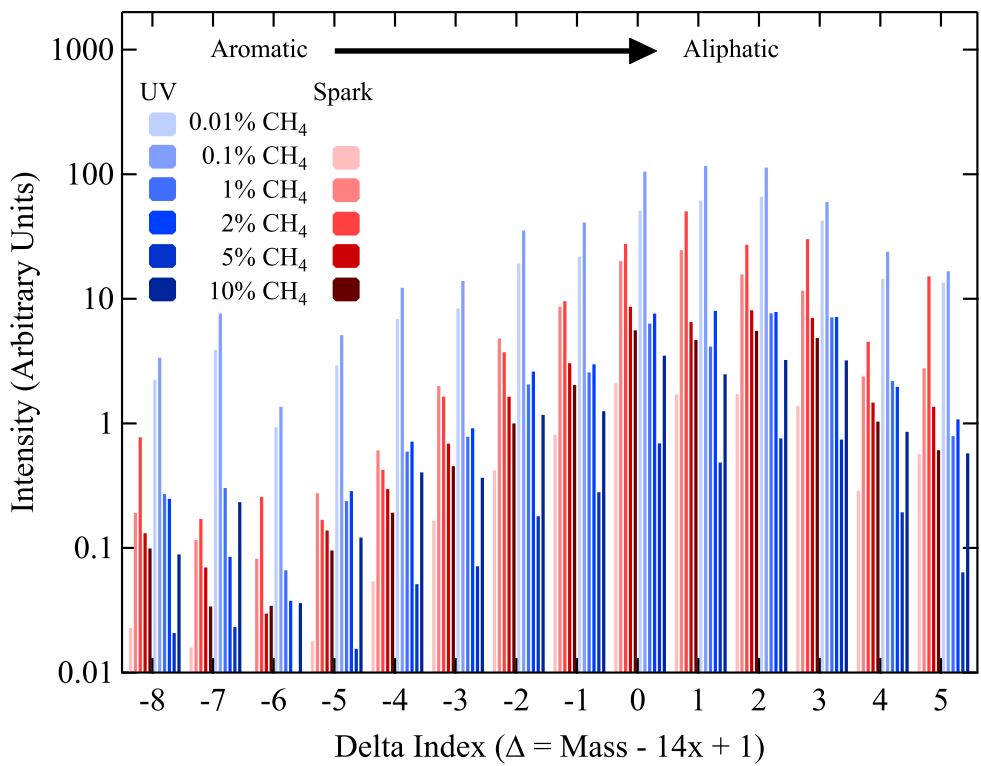}}
\caption{Shown here is the abundance distribution as a function of delta index ($\Delta =$ Mass$- 14x$+1) for the spark (red) and UV (blue) experiments for a range of initial CH$_{4}$ concentrations from the AMS measurements of the particle phase products. The data presented here were obtained using the standard flow rate ($\sim$130 sccm). Positive delta indices correspond to a greater degree of saturation in the molecules.  \label{fig:delta_ams}}
\end{figure}

The AMS has higher mass resolution than the PIT-MS, allowing for definitive ion identification for spectral peaks where it is not possible in the PIT-MS data. Additionally, the PIT-MS uses a soft ionization source, unlike the AMS. These differences make direct comparison between the AMS and PIT-MS measurements more difficult. However, similar techniques can be used by binning the AMS measurements to the same resolution as the PIT-MS measurements and accounting for the mass shift resulting from the PIT-MS proton addition. Figure \ref{fig:delta_ams} shows the delta index for the AMS mass spectra. The shape is similar for all gas mixtures investigated and for both energy sources, which is not particularly surprising given the similarities between the 2\% CH$_{4}$ mass spectra shown in Figure \ref{fig:ams2per}. Interestingly, the AMS measurements have higher (and therefore more aliphatic) average delta indices than were observed in the PIT-MS measurements of the gas phase products. This difference has at least two possible explanations. First, as discussed previously, the PIT-MS measurements are biased against small hydrocarbons, with the strongest bias against the most aliphatic species (alkanes). It is therefore possible that inclusion of these species would shift the value of the average delta index into agreement with the AMS measurements of the particles. Second, as pointed out by \citet{Imanaka:2010} it is possible that the particles contain more hydrogen (a higher degree of saturation) than the gases from which they formed since \citet{Sekine:2008} showed that hydrogenation is more efficient than H abstraction. Due to the limitations of the data, we cannot distinguish those two possibilities in our experiments. However, both the gas phase and aerosol products are composed primarily of molecules with a low degree of aromaticity. 

\begin{figure}
\centering
\resizebox{5.5in}{!}
{\includegraphics{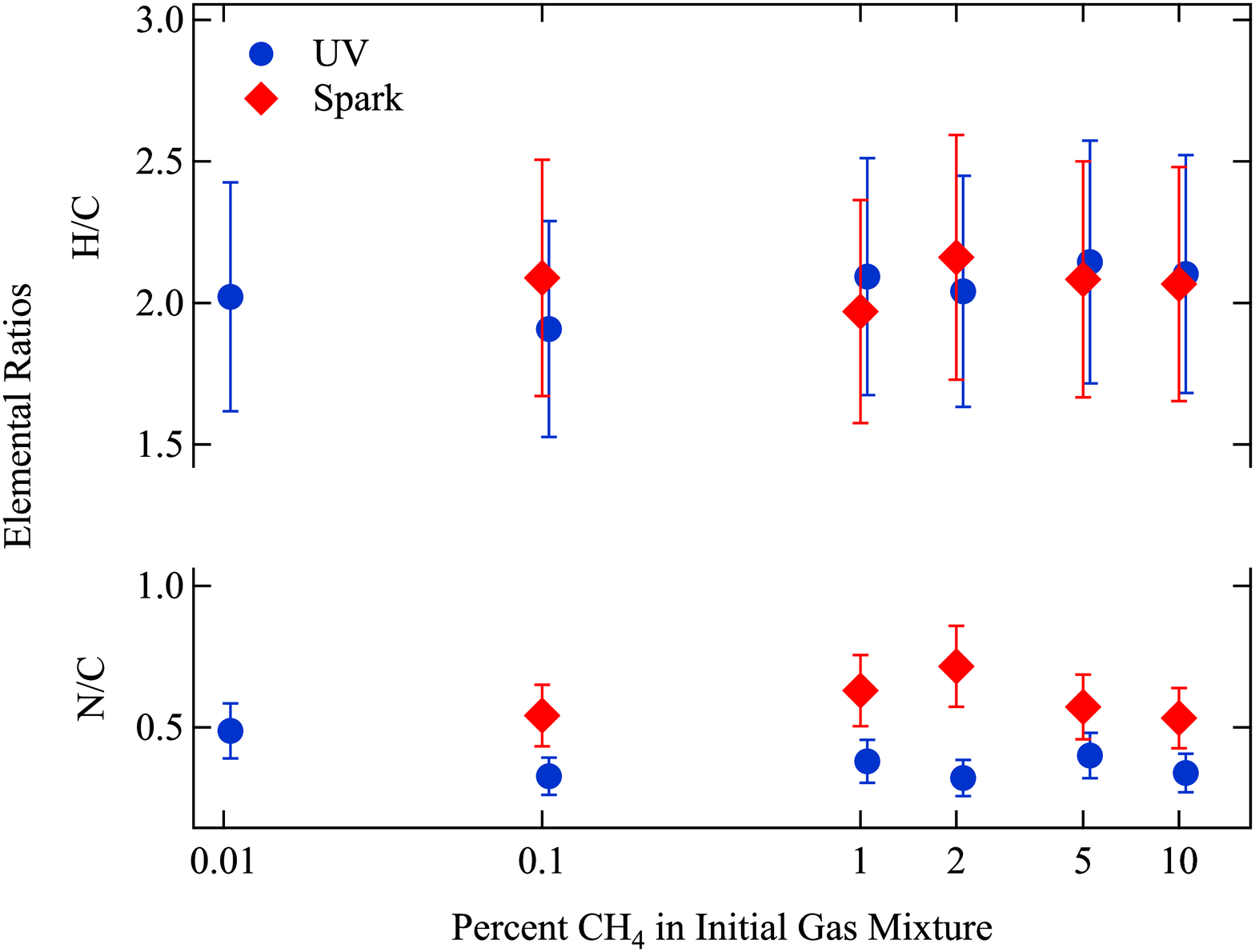}}
\caption{Measured H/C and N/C in the particle phase as a function of initial CH$_{4}$ concentration for both the spark and UV energy sources. The 0.01\% spark experiment did not produce enough aerosol for reliable elemental composition measurements. Here 20\% error bars were used based on the work of \citet{Aiken:2007}. The measured errors are significantly smaller. For clarity, the UV measurements are offset slightly on the x axis. \label{fig:element}}
\end{figure}

\subsubsection{Particle nitrogen incorporation}

The degree of nitrogen incorporation is much better constrained for the aerosol using the AMS measurements. Figure \ref{fig:element} shows the elemental ratios, H/C and N/C, in the aerosol as a function of initial CH$_{4}$ concentration for both the spark and UV experiments. Within the error bars, the H/C ratios are the same, slightly greater than 2, for every experiment indicating that the H/C ratio is independent of CH$_{4}$ concentration and energy source. However, for all initial CH$_{4}$ concentrations the spark samples have a higher N/C ratio than the UV samples. On average the spark tholins are 44$\pm$6\% nitrogen by mass, while the UV tholins are 23$\pm$3\% nitrogen by mass. This is consistent with the information obtained from the gas phase composition measurements and is not surprising given that the spark energy source is more energetic and capable of direct dissociation of N$_{2}$. Interestingly, the N/C ratio for the UV experiments appears to be independent of CH$_{4}$ concentration, while the N/C ratio for the spark experiments peaks at 2\% CH$_{4}$. Further, the shape of the N/C curve is similar to the shape of the aerosol production curve for the same measurements \citep{Horst:2013}, which indicates that nitrogen may play an important role in determining the amount of aerosol produced. The highest N/C value for the UV samples is observed in samples produced from 0.01\% CH$_{4}$ which is also the region where the highest aerosol production occurs, though there does not seem to be an overall trend for the UV experiments. 

\begin{figure}
\centering
\resizebox{5.5in}{!}
{\includegraphics{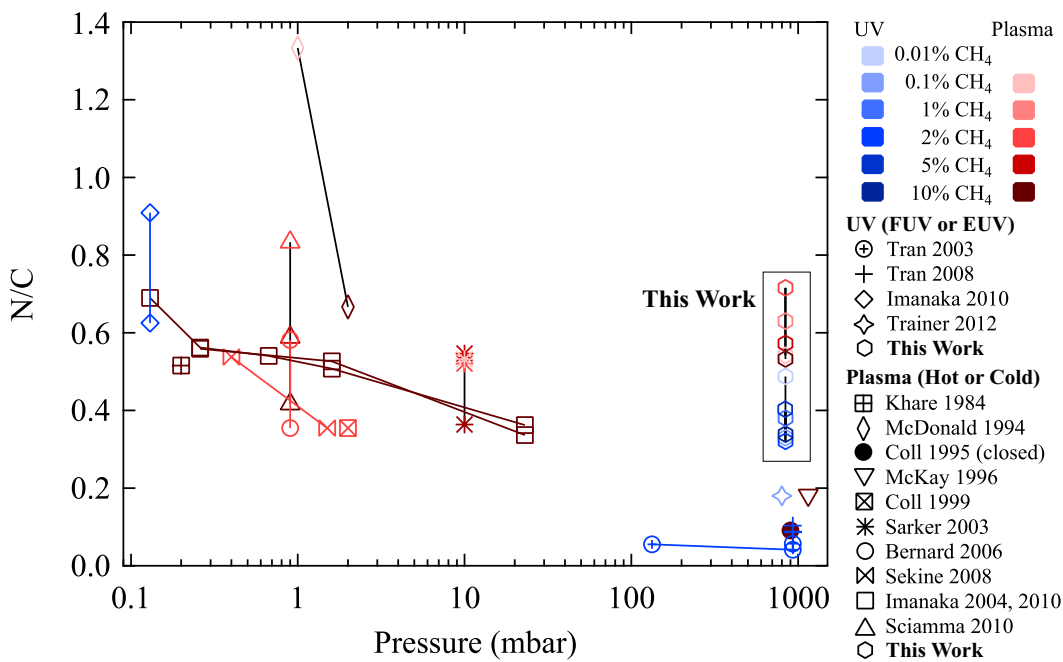}}
\caption{Comparison of reported N/C ratios for tholins produced from a number of different experiments as a function of pressure. UV (all wavelengths) experiments are shown in blue and plasma (both hot and cold) are shown in red. Points connected by lines come from the same work and indicate a study that varied either pressure or initial gas composition, except for the \citet{Bernard:2006} points, which are samples taken from two different distances from the inlet in the experiment. Due to the use of relatively low energy photons from Hg lamp (185/254 nm), the experiments of \citet{Tran:2003}, \citet{Tran:2003b}, and \citet{Tran:2008} used gas mixtures that included additional species. \citet{Tran:2003} and \citet{Tran:2003b}  0.2\% H$_{2}$, 0.035\% C$_{2}$H$_{2}$, 0.03\% C$_{2}$H$_{4}$, and 0.0017\% HC$_{3}$N. \citet{Tran:2008} gas mixtures included 0.2\% H$_{2}$, 0.04\% C$_{2}$H$_{2}$, 0.03\% C$_{2}$H$_{4}$, and 0.002\% HC$_{3}$N and one experiment with 0.02\% HCN. 
\label{fig:nc_pressure}}
\end{figure}

\nocite{Imanaka:2004, Khare:1984, Sciamma:2010, Sarker:2003, McDonald:1994}


Elemental composition is one of the most frequently measured parameters for tholin and as such we are able to put our measurements into context with other experiments. Figure \ref{fig:nc_pressure} plots N/C as a function of pressure for numerous tholin experiments. It is important to remember that there are a number of differences between the experiments including gas composition (including variations in percent methane and addition of other trace gases such as C$_{2}$H$_{2}$), energy source, and flow rate. The elemental composition was also obtained using different methods depending on the experiment. Despite these differences, the vast majority of experiments find N/C ratios between about 0.35 and 0.9. With only one exception, all of the experiments find that there is more carbon in the particles than nitrogen. Most of the outliers with lower values come from experiments using photons from a Hg lamp, which are even lower energy than the photons used in this experiment. Although no single experiment has completely spanned the range of both pressure and initial gas composition presented here, taken collectively the experiments seem to indicate that the initial CH$_{4}$ concentration is of comparable importance to the experimental pressure in determining the N/C ratio of the particles. This idea reinforced by the experiments that have varied CH$_{4}$ concentration at a single pressure and see large variations in the N/C ratio \citep{Sarker:2003, Bernard:2006, Sciamma:2010}.

\begin{figure}
\centering
\resizebox{5.5in}{!}
{\includegraphics{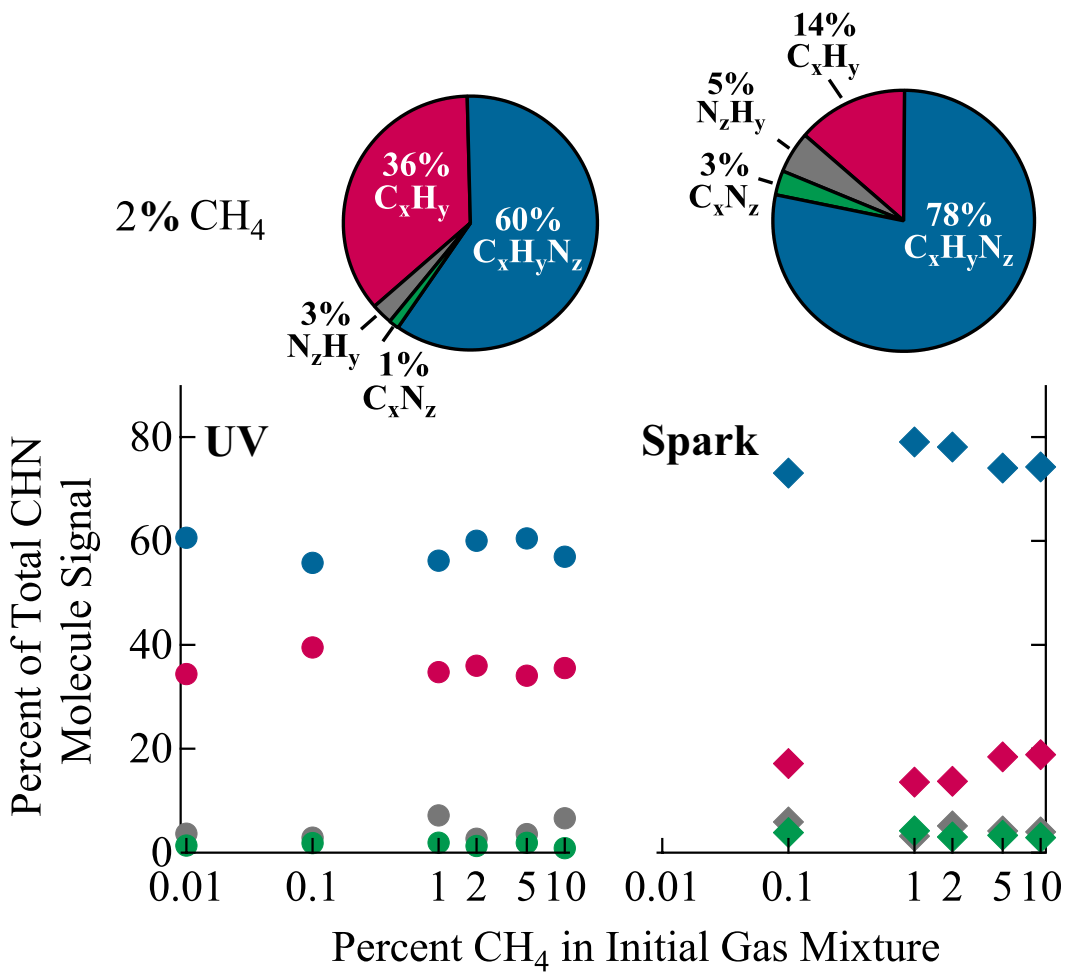}}
\caption{Shown here is the distribution of signal among different CHN molecule groups as a function of initial CH$_{4}$ concentration for the UV (left) and spark (right) produced particles from the AMS measurements. The pie charts (top) compare the results for 2\% CH$_{4}$ for the UV and spark experiments.  \label{fig:nitrogen_pie}}
\end{figure}

Due to the fragmentation of the aerosol molecules in the AMS, it is more difficult to look at the behavior of functional groups. However, as shown in Figure \ref{fig:nitrogen_pie} we can divide the ions into different CHN groups, such as C$_{x}$H$_{y}$ and C$_{x}$H$_{y}$N$_{z}$, and look at their behavior as a function of CH$_{4}$ concentration and energy source. For 2\% CH$_{4}$ samples, both the UV and spark are comprised mostly of ions in the C$_{x}$H$_{y}$N$_{z}$ group. However, the UV contains relatively more C$_{x}$H$_{y}$ ions, mostly at the expense of C$_{x}$H$_{y}$N$_{z}$. The C$_{x}$N$_{z}$ and N$_{z}$H$_{y}$ groups totaled less than 10\% of the signal for both samples. In general, the CHN group partitioning does not strongly depend on the initial CH$_{4}$ concentration, which is consistent with the elemental ratios; although the previously mentioned increase in nitrogen relative to carbon for the spark samples at 2\% CH$_{4}$ results mainly from a shift between abundances in the C$_{x}$H$_{y}$ and C$_{x}$H$_{y}$N$_{z}$ groups. 

As detailed in \citet{Cable:2014}, UV and spark samples produced using the same experimental setup were collected for further analysis using microchip nonaqueous capillary electrophoresis. This technique provided quantitative measurements of the primary amine content of 2\% CH$_{4}$ spark tholin samples and a 0.1\% CH$_{4}$ UV tholin sample. Primary amines up to C24 were measured in both the spark and UV samples. The spark samples contained $\sim$2-3 times greater total primary amine content (by weight percent) than the UV sample. In general the primary amine content was relatively low,  0.3-0.6\% for the spark sample and less than 0.2\% for the UV sample \citep{Cable:2014}. This analysis when combined with the AMS measurements indicates that the nitrogen in the particles is most likely found in nitriles rather than amines. This interpretation is consistent with the gas phase measurements which show that nitriles are the most abundant nitrogen containing functional group in the gas phase. Nitriles are also the most abundant nitrogen functional group detected in the gas phase on Titan (see Table \ref{table:comp}).

\section{Summary}
Through the use of three complementary measurement instruments (high-resolution time-of-flight aerosol mass spectrometer, proton-transfer ion-trap mass spectrometer, and a scanning mobility particle sizer) we characterize the gas and solid products of chemistry initiated by FUV photons or spark discharge in gas mixtures ranging from 0.01\% CH$_{4}$/99.99\% N$_{2}$ to 10\% CH$_{4}$/90\% N$_{2}$.

Here we summarize the main conclusions of this work:

\begin{itemize}

\item To our knowledge, we have measured the heaviest gas phase products produced in a Titan atmosphere simulation experiment.
\item We definitively demonstrate the presence of nitrogen incorporation in the gas phase products. Since the mechanism responsible for nitrogen incorporation in these experiments is not yet understood, we do not yet know if the mechanism is important in Titan's atmosphere. However, it is possible that an important reaction or set of reactions is missing from current models of Titan's atmosphere
\item The UV experiments reproduce the absolute abundances measured in Titan's stratosphere for a number of gas phase species including C$_{4}$H$_{2}$, C$_{6}$H$_{6}$, HCN, CH$_{3}$CN, HC$_{3}$N, and C$_{2}$H$_{5}$CN.
\item The particles appear to be more aliphatic than the gas phase products, consistent with the suggestion of \citet{Imanaka:2010} that particles contain more hydrogen (a higher degree of saturation) than the gases from which they formed since \citet{Sekine:2008} showed that hydrogenation is more efficient than H abstraction. 
\item The particle composition measurements, in combination with \citet{Cable:2014}, indicate that the nitrogen is most likely contained in nitriles, not amines, which is consistent with the gas phase measurements that show that nitrogen bearing species abundance decreases with increasing saturation
\item Within the limitations of our measurements, discussed at length earlier, the nitrogen content of the gas phase products generally decreases with increasing CH$_{4}$ concentration. 
\item In the particles, the H/C ratio is not strongly dependent on gas mixture. There is no obvious trend in the N/C ratio with changing CH$_{4}$ concentration. However, the N/C ratio does vary as a function of methane concentration for the spark experiments with the highest N/C ratio measured in the 2\% CH$_{4}$ sample.

\end{itemize}

\section{Acknowledgements}

SMH was supported in part by NSF Astronomy and Astrophysics Postdoctoral Fellowship AST-1102827. This work was also supported by NASA grant NNX11AD82G.









\label{}

\clearpage

\bibliographystyle{elsarticle-harv}
\bibliography{Titanoxygen}

\begin{thebibliography}{84}
\expandafter\ifx\csname natexlab\endcsname\relax\def\natexlab#1{#1}\fi
\expandafter\ifx\csname url\endcsname\relax
  \def\url#1{\texttt{#1}}\fi
\expandafter\ifx\csname urlprefix\endcsname\relax\def\urlprefix{URL }\fi

\bibitem[{Aiken et~al.({2007})Aiken, DeCarlo, and Jimenez}]{Aiken:2007}
Aiken, A.~C., DeCarlo, P.~F., Jimenez, J.~L., {NOV 1} {2007}. {Elemental
  analysis of organic species with electron ionization high-resolution mass
  spectrometry}. {Analytical Chemistry} {79}~({21}), {8350--8358}.

\bibitem[{Aiken et~al.({2008})Aiken, Decarlo, Kroll, Worsnop, Huffman,
  Docherty, Ulbrich, Mohr, Kimmel, Sueper, Sun, Zhang, Trimborn, Northway,
  Ziemann, Canagaratna, Onasch, Alfarra, Prevot, Dommen, Duplissy, Metzger,
  Baltensperger, and Jimenez}]{Aiken:2008}
Aiken, A.~C., Decarlo, P.~F., Kroll, J.~H., Worsnop, D.~R., Huffman, J.~A.,
  Docherty, K.~S., Ulbrich, I.~M., Mohr, C., Kimmel, J.~R., Sueper, D., Sun,
  Y., Zhang, Q., Trimborn, A., Northway, M., Ziemann, P.~J., Canagaratna,
  M.~R., Onasch, T.~B., Alfarra, M.~R., Prevot, A. S.~H., Dommen, J., Duplissy,
  J., Metzger, A., Baltensperger, U., Jimenez, J.~L., {JUN 15} {2008}. {O/C and
  OM/OC ratios of primary, secondary, and ambient organic aerosols with
  high-resolution time-of-flight aerosol mass spectrometry}. {Environmental
  Science \& Technology} {42}~({12}), {4478--4485}.

\bibitem[{{Bernard} et~al.(2006){Bernard}, {Quirico}, {Brissaud}, {Montagnac},
  {Reynard}, {McMillan}, {Coll}, {Nguyen}, {Raulin}, and
  {Schmitt}}]{Bernard:2006}
{Bernard}, J.-M., {Quirico}, E., {Brissaud}, O., {Montagnac}, G., {Reynard},
  B., {McMillan}, P., {Coll}, P., {Nguyen}, M., {Raulin}, F., {Schmitt}, B.,
  Nov. 2006. {Reflectance spectra and chemical structure of Titan's tholins:
  Application to the analysis of Cassini Huygens observations}. Icarus 185,
  301--307.

\bibitem[{{Borucki} et~al.(1988){Borucki}, {Giver}, {McKay}, {Scattergood}, and
  {Parris}}]{Borucki:1988}
{Borucki}, W.~J., {Giver}, L.~P., {McKay}, C.~P., {Scattergood}, T., {Parris},
  J.~E., Oct. 1988. {Lightning production of hydrocarbons and HCN on Titan -
  Laboratory measurements}. Icarus 76, 125--134.

\bibitem[{{Cable} et~al.(2014){Cable}, {H{\"o}rst}, {He}, {Stockton}, {Mora},
  {Tolbert}, {Smith}, and {Willis}}]{Cable:2014}
{Cable}, M.~L., {H{\"o}rst}, S.~M., {He}, C., {Stockton}, A.~M., {Mora}, M.~F.,
  {Tolbert}, M.~A., {Smith}, M.~A., {Willis}, P.~A., Oct. 2014. {Identification
  of primary amines in Titan tholins using microchip nonaqueous capillary
  electrophoresis}. Earth and Planetary Science Letters 403, 99--107.

\bibitem[{Cable et~al.(2012)Cable, H\"orst, Hodyss, Beauchamp, Smith, and
  Willis}]{Cable:2012}
Cable, M.~L., H\"orst, S.~M., Hodyss, R., Beauchamp, P.~M., Smith, M.~A.,
  Willis, P.~A., 2012. Titan tholins: Simulating titan organic chemistry in the
  cassini-huygens era. Chemical Reviews 112~(3), 1882--1909.

\bibitem[{{Carrasco} et~al.(2012){Carrasco}, {Gautier}, {Es-sebbar}, {Pernot},
  and {Cernogora}}]{Carrasco:2012}
{Carrasco}, N., {Gautier}, T., {Es-sebbar}, E.-t., {Pernot}, P., {Cernogora},
  G., May 2012. {Volatile products controlling Titan's tholins production}.
  Icarus 219, 230--240.

\bibitem[{Chan et~al.({1993})Chan, Cooper, Sodhi, and Brion}]{Chan:1993d}
Chan, W., Cooper, G., Sodhi, R., Brion, C., {FEB 15} {1993}. {Absolute optical
  oscillator-strengths for discrete and continuum photoabsorption of molecular
  nitrogen (11-200 eV)}. {Chemical Physics} {170}~({1}), {81--97}.

\bibitem[{Chen and Wu({2004})}]{Chen:2004}
Chen, F., Wu, C., {MAY 1} {2004}. {Temperature-dependent photoabsorption cross
  sections in the VUV-UV region. I. Methane and ethane}. {Journal of
  Quantitative Spectroscopy and Radiative Transfer} {85}~({2}), {195--209}.

\bibitem[{Chhabra et~al.({2010})Chhabra, Flagan, and Seinfeld}]{Chhabra:2010}
Chhabra, P.~S., Flagan, R.~C., Seinfeld, J.~H., {2010}. {Elemental analysis of
  chamber organic aerosol using an aerodyne high-resolution aerosol mass
  spectrometer}. {Atmospheric Chemistry and Physics} {10}~({9}), {4111--4131}.

\bibitem[{{Coates} et~al.(2007){Coates}, {Crary}, {Lewis}, {Young}, {Waite},
  and {Sittler}}]{Coates:2007}
{Coates}, A.~J., {Crary}, F.~J., {Lewis}, G.~R., {Young}, D.~T., {Waite},
  J.~H., {Sittler}, E.~C., Nov. 2007. {Discovery of heavy negative ions in
  Titan's ionosphere}. Geophys. Res. Lett. 34, 22103--+.

\bibitem[{{Coll} et~al.(1995){Coll}, {Coscia}, {Gazeau}, {de Vanssay},
  {Guillemin}, and {Raulin}}]{Coll:1995}
{Coll}, P., {Coscia}, D., {Gazeau}, M.~C., {de Vanssay}, E., {Guillemin},
  J.~C., {Raulin}, F., Aug. 1995. {Organic chemistry in Titan's atmosphere: new
  data from laboratory simulations at low temperature}. Adv. Space Res. 16,
  93--103.

\bibitem[{{Coll} et~al.(1999){Coll}, {Coscia}, {Smith}, {Gazeau},
  {Ram{\'{\i}}rez}, {Cernogora}, {Isra{\"e}l}, and {Raulin}}]{Coll:1999}
{Coll}, P., {Coscia}, D., {Smith}, N., {Gazeau}, M., {Ram{\'{\i}}rez}, S.~I.,
  {Cernogora}, G., {Isra{\"e}l}, G., {Raulin}, F., Oct. 1999. {Experimental
  laboratory simulation of Titan's atmosphere: aerosols and gas phase}. Planet.
  Space Sci. 47, 1331--1340.

\bibitem[{{Coll} et~al.(2013){Coll}, {Navarro-Gonz{\'a}lez}, {Szopa}, {Poch},
  {Ram{\'{\i}}rez}, {Coscia}, {Raulin}, {Cabane}, {Buch}, and
  {Isra{\"e}l}}]{Coll:2013}
{Coll}, P., {Navarro-Gonz{\'a}lez}, R., {Szopa}, C., {Poch}, O.,
  {Ram{\'{\i}}rez}, S.~I., {Coscia}, D., {Raulin}, F., {Cabane}, M., {Buch},
  A., {Isra{\"e}l}, G., Mar. 2013. {Can laboratory tholins mimic the chemistry
  producing Titan's aerosols? A review in light of ACP experimental results}.
  Planet. Space Sci. 77, 91--103.

\bibitem[{{Cordiner} et~al.(2015){Cordiner}, {Palmer}, {Nixon}, {Irwin},
  {Teanby}, {Charnley}, {Mumma}, {Kisiel}, {Serigano}, {Kuan}, {Chuang}, and
  {Wang}}]{Cordiner:2015}
{Cordiner}, M.~A., {Palmer}, M.~Y., {Nixon}, C.~A., {Irwin}, P.~G.~J.,
  {Teanby}, N.~A., {Charnley}, S.~B., {Mumma}, M.~J., {Kisiel}, Z., {Serigano},
  J., {Kuan}, Y.-J., {Chuang}, Y.-L., {Wang}, K.-S., Feb. 2015. {Ethyl Cyanide
  On Titan: Spectroscopic Detection and Mapping Using Alma}. ApJL 800, L14.

\bibitem[{{Courtin} et~al.(2007){Courtin}, {Sim}, {Kim}, and
  {Gautier}}]{Courtin:2007}
{Courtin}, R.~D., {Sim}, C.~K., {Kim}, S.~J., {Gautier}, D., Oct. 2007. {The
  Tropospheric Abundance of H$_{2}$ on Titan from the Cassini CIRS
  Investigation}. In: BAAS. Vol.~38 of Bulletin of the American Astronomical
  Society. pp. 529--+.

\bibitem[{{Coustenis} et~al.(2007){Coustenis}, {Achterberg}, {Conrath},
  {Jennings}, {Marten}, {Gautier}, {Nixon}, {Flasar}, {Teanby}, {B{\'e}zard},
  {Samuelson}, {Carlson}, {Lellouch}, {Bjoraker}, {Romani}, {Taylor}, {Irwin},
  {Fouchet}, {Hubert}, {Orton}, {Kunde}, {Vinatier}, {Mondellini}, {Abbas}, and
  {Courtin}}]{Coustenis:2007}
{Coustenis}, A., {Achterberg}, R.~K., {Conrath}, B.~J., {Jennings}, D.~E.,
  {Marten}, A., {Gautier}, D., {Nixon}, C.~A., {Flasar}, F.~M., {Teanby},
  N.~A., {B{\'e}zard}, B., {Samuelson}, R.~E., {Carlson}, R.~C., {Lellouch},
  E., {Bjoraker}, G.~L., {Romani}, P.~N., {Taylor}, F.~W., {Irwin}, P.~G.~J.,
  {Fouchet}, T., {Hubert}, A., {Orton}, G.~S., {Kunde}, V.~G., {Vinatier}, S.,
  {Mondellini}, J., {Abbas}, M.~M., {Courtin}, R., Jul. 2007. {The composition
  of Titan's stratosphere from Cassini/CIRS mid-infrared spectra}. Icarus 189,
  35--62.

\bibitem[{{Coustenis} et~al.(2010){Coustenis}, {Jennings}, {Nixon},
  {Achterberg}, {Lavvas}, {Vinatier}, {Teanby}, {Bjoraker}, {Carlson}, {Piani},
  {Bampasidis}, {Flasar}, and {Romani}}]{Coustenis:2010}
{Coustenis}, A., {Jennings}, D.~E., {Nixon}, C.~A., {Achterberg}, R.~K.,
  {Lavvas}, P., {Vinatier}, S., {Teanby}, N.~A., {Bjoraker}, G.~L., {Carlson},
  R.~C., {Piani}, L., {Bampasidis}, G., {Flasar}, F.~M., {Romani}, P.~N., May
  2010. {Titan trace gaseous composition from CIRS at the end of the
  Cassini-Huygens prime mission}. Icarus 207, 461--476.

\bibitem[{{Crary} et~al.(2009){Crary}, {Magee}, {Mandt}, {Waite}, {Westlake},
  and {Young}}]{Crary:2009}
{Crary}, F.~J., {Magee}, B.~A., {Mandt}, K., {Waite}, J.~H., {Westlake}, J.,
  {Young}, D.~T., Dec. 2009. {Heavy ions, temperatures and winds in Titan's
  ionosphere: Combined Cassini CAPS and INMS observations}. Planet. Space Sci.
  57, 1847--1856.

\bibitem[{{Cui} et~al.(2009){Cui}, {Yelle}, {Vuitton}, {Waite}, {Kasprzak},
  {Gell}, {Niemann}, {M{\"u}ller-Wodarg}, {Borggren}, {Fletcher}, {Patrick},
  {Raaen}, and {Magee}}]{Cui:2009}
{Cui}, J., {Yelle}, R.~V., {Vuitton}, V., {Waite}, J.~H., {Kasprzak}, W.~T.,
  {Gell}, D.~A., {Niemann}, H.~B., {M{\"u}ller-Wodarg}, I.~C.~F., {Borggren},
  N., {Fletcher}, G.~G., {Patrick}, E.~L., {Raaen}, E., {Magee}, B.~A., Apr.
  2009. {Analysis of Titan's neutral upper atmosphere from Cassini Ion Neutral
  Mass Spectrometer measurements}. Icarus 200, 581--615.

\bibitem[{de~Gouw and Warneke(2007)}]{deGouw:2007}
de~Gouw, J., Warneke, C., 2007. Measurements of volatile organic compounds in
  the earth's atmosphere using proton-transfer-reaction mass spectrometry. Mass
  Spectrometry Reviews 26~(2), 223--257.

\bibitem[{DeCarlo et~al.({2006})DeCarlo, Kimmel, Trimborn, Northway, Jayne,
  Aiken, Gonin, Fuhrer, Horvath, Docherty, Worsnop, and Jimenez}]{decarlo:2006}
DeCarlo, P.~F., Kimmel, J.~R., Trimborn, A., Northway, M.~J., Jayne, J.~T.,
  Aiken, A.~C., Gonin, M., Fuhrer, K., Horvath, T., Docherty, K.~S., Worsnop,
  D.~R., Jimenez, J.~L., {DEC 15} {2006}. {Field-deployable, high-resolution,
  time-of-flight aerosol mass spectrometer}. {Analytical Chemistry}
  {78}~({24}), {8281--8289}.

\bibitem[{{Delpech}(1994)}]{Delpech:1994}
{Delpech}, C., Jun. 1994. {Infrared spectra of triacetylene in the 4000-220
  cm-1 region: Absolute band intensity and implications for the atmosphere of
  Titan}. Spectrochimica Acta Part A: Molecular Spectroscopy 50, 1095--1100.

\bibitem[{Dodonova(1966)}]{Dodonova:1966}
Dodonova, N.~Y., 1966. Activation of nitrogen by vacuum ultraviolet radiation.
  Russian Journal of Physical Chemistry, USSR 40~(5), 523.

\bibitem[{Drewnick et~al.({2005})Drewnick, Hings, DeCarlo, Jayne, Gonin,
  Fuhrer, Weimer, Jimenez, Demerjian, Borrmann, and Worsnop}]{Drewnick:2005}
Drewnick, F., Hings, S., DeCarlo, P., Jayne, J., Gonin, M., Fuhrer, K., Weimer,
  S., Jimenez, J., Demerjian, K., Borrmann, S., Worsnop, D., {JUL} {2005}. {A
  new time-of-flight aerosol mass spectrometer (TOF-AMS) - Instrument
  description and first field deployment}. {Aerosol Science and Technology}
  {39}~({7}), {637--658}.

\bibitem[{{Flasar} et~al.(2005){Flasar}, {Achterberg}, {Conrath}, {Gierasch},
  {Kunde}, {Nixon}, {Bjoraker}, {Jennings}, {Romani}, {Simon-Miller},
  {B{\'e}zard}, {Coustenis}, {Irwin}, {Teanby}, {Brasunas}, {Pearl}, {Segura},
  {Carlson}, {Mamoutkine}, {Schinder}, {Barucci}, {Courtin}, {Fouchet},
  {Gautier}, {Lellouch}, {Marten}, {Prang{\'e}}, {Vinatier}, {Strobel},
  {Calcutt}, {Read}, {Taylor}, {Bowles}, {Samuelson}, {Orton}, {Spilker},
  {Owen}, {Spencer}, {Showalter}, {Ferrari}, {Abbas}, {Raulin}, {Edgington},
  {Ade}, and {Wishnow}}]{Flasar:2005}
{Flasar}, F.~M., {Achterberg}, R.~K., {Conrath}, B.~J., {Gierasch}, P.~J.,
  {Kunde}, V.~G., {Nixon}, C.~A., {Bjoraker}, G.~L., {Jennings}, D.~E.,
  {Romani}, P.~N., {Simon-Miller}, A.~A., {B{\'e}zard}, B., {Coustenis}, A.,
  {Irwin}, P.~G.~J., {Teanby}, N.~A., {Brasunas}, J., {Pearl}, J.~C., {Segura},
  M.~E., {Carlson}, R.~C., {Mamoutkine}, A., {Schinder}, P.~J., {Barucci}, A.,
  {Courtin}, R., {Fouchet}, T., {Gautier}, D., {Lellouch}, E., {Marten}, A.,
  {Prang{\'e}}, R., {Vinatier}, S., {Strobel}, D.~F., {Calcutt}, S.~B., {Read},
  P.~L., {Taylor}, F.~W., {Bowles}, N., {Samuelson}, R.~E., {Orton}, G.~S.,
  {Spilker}, L.~J., {Owen}, T.~C., {Spencer}, J.~R., {Showalter}, M.~R.,
  {Ferrari}, C., {Abbas}, M.~M., {Raulin}, F., {Edgington}, S., {Ade}, P.,
  {Wishnow}, E.~H., May 2005. {Titan's Atmospheric Temperatures, Winds, and
  Composition}. Science 308, 975--978.

\bibitem[{Fujii and Arai(1999)}]{Fujii:1999}
Fujii, T., Arai, N., 1999. {Analysis of N-containing hydrocarbon species
  produced by a CH$_{4}$/N$_{2}$ microwave discharge: Simulation of Titan's
  atmosphere}. The Astrophysical Journal 519~(2), 858.

\bibitem[{{Gautier} et~al.(2011){Gautier}, {Carrasco}, {Buch}, {Szopa},
  {Sciamma-O'Brien}, and {Cernogora}}]{Gautier:2011}
{Gautier}, T., {Carrasco}, N., {Buch}, A., {Szopa}, C., {Sciamma-O'Brien}, E.,
  {Cernogora}, G., Jun. 2011. {Nitrile gas chemistry in Titan's atmosphere}.
  Icarus 213, 625--635.

\bibitem[{{Gupta} et~al.(1981){Gupta}, {Ochiai}, and
  {Ponnamperuma}}]{Gupta:1981}
{Gupta}, S.~K., {Ochiai}, E., {Ponnamperuma}, C., Oct. 1981. {Organic synthesis
  in the atmosphere of Titan}. Nature 293, 725--727.

\bibitem[{{Hodyss} et~al.(2011){Hodyss}, {Howard}, {Johnson}, {Goguen}, and
  {Kanik}}]{Hodyss:2011}
{Hodyss}, R., {Howard}, H.~R., {Johnson}, P.~V., {Goguen}, J.~D., {Kanik}, I.,
  Aug. 2011. {Formation of radical species in photolyzed CH$_{4}$:N$_{2}$
  ices}. Icarus 214, 748--753.

\bibitem[{{H{\"o}rst}(2017)}]{Horst:2017}
{H{\"o}rst}, S.~M., Mar. 2017. {Titan's atmosphere and climate}. Journal of
  Geophysical Research (Planets) 122, 432--482.

\bibitem[{{H{\"o}rst} and {Tolbert}(2013)}]{Horst:2013}
{H{\"o}rst}, S.~M., {Tolbert}, M.~A., Jun. 2013. {In Situ Measurements of the
  Size and Density of Titan Aerosol Analogs}. ApJL 770, L10.

\bibitem[{{H{\"o}rst} et~al.(2012){H{\"o}rst}, {Yelle}, {Buch}, {Carrasco},
  {Cernogora}, {Dutuit}, {Quirico}, {Sciamma-O'Brien}, {Smith}, {Somogyi},
  {Szopa}, {Thissen}, and {Vuitton}}]{Horst:2012}
{H{\"o}rst}, S.~M., {Yelle}, R.~V., {Buch}, A., {Carrasco}, N., {Cernogora},
  G., {Dutuit}, O., {Quirico}, E., {Sciamma-O'Brien}, E., {Smith}, M.~A.,
  {Somogyi}, {\'A}., {Szopa}, C., {Thissen}, R., {Vuitton}, V., Sep. 2012.
  {Formation of Amino Acids and Nucleotide Bases in a Titan Atmosphere
  Simulation Experiment}. Astrobiology 12, 809--817.

\bibitem[{Hunter and Lias({1998})}]{Hunter:1998}
Hunter, E., Lias, S., {MAY-JUN} {1998}. {Evaluated gas phase basicities and
  proton affinities of molecules: An update}. {Journal of Physical and Chemical
  Reference Data} {27}~({3}), {413--656}.

\bibitem[{{Imanaka} et~al.(2004){Imanaka}, {Khare}, {Elsila}, {Bakes}, {McKay},
  {Cruikshank}, {Sugita}, {Matsui}, and {Zare}}]{Imanaka:2004}
{Imanaka}, H., {Khare}, B.~N., {Elsila}, J.~E., {Bakes}, E.~L.~O., {McKay},
  C.~P., {Cruikshank}, D.~P., {Sugita}, S., {Matsui}, T., {Zare}, R.~N., Apr.
  2004. {Laboratory experiments of Titan tholin formed in cold plasma at
  various pressures: implications for nitrogen-containing polycyclic aromatic
  compounds in Titan haze}. Icarus 168, 344--366.

\bibitem[{{Imanaka} and {Smith}(2007)}]{Imanaka:2007}
{Imanaka}, H., {Smith}, M.~A., Jan. 2007. {Role of photoionization in the
  formation of complex organic molecules in Titan's upper atmosphere}. Geophys.
  Res. Lett. 34, 2204--+.

\bibitem[{{Imanaka} and {Smith}(2009)}]{Imanaka:2009}
{Imanaka}, H., {Smith}, M.~A., Aug. 2009. {EUV Photochemical Production of
  Unsaturated Hydrocarbons: Implications to EUV Photochemistry in Titan and
  Jovian Planets}. J. Phys. Chem. A 42, 11187--11194.

\bibitem[{{Imanaka} and {Smith}(2010)}]{Imanaka:2010}
{Imanaka}, H., {Smith}, M.~A., Jul. 2010. {Formation of nitrogenated organic
  aerosols in the Titan Upper Atmosphere}. PNAS 28, 12423--12428.

\bibitem[{{Isra{\"e}l} et~al.(2005){Isra{\"e}l}, {Szopa}, {Raulin}, {Cabane},
  {Niemann}, {Atreya}, {Bauer}, {Brun}, {Chassefi{\`e}re}, {Coll}, {Cond{\'e}},
  {Coscia}, {Hauchecorne}, {Millian}, {Nguyen}, {Owen}, {Riedler}, {Samuelson},
  {Siguier}, {Steller}, {Sternberg}, and {Vidal-Madjar}}]{Israel:2005}
{Isra{\"e}l}, G., {Szopa}, C., {Raulin}, F., {Cabane}, M., {Niemann}, H.~B.,
  {Atreya}, S.~K., {Bauer}, S.~J., {Brun}, J., {Chassefi{\`e}re}, E., {Coll},
  P., {Cond{\'e}}, E., {Coscia}, D., {Hauchecorne}, A., {Millian}, P.,
  {Nguyen}, M., {Owen}, T., {Riedler}, W., {Samuelson}, R.~E., {Siguier}, J.,
  {Steller}, M., {Sternberg}, R., {Vidal-Madjar}, C., Dec. 2005. {Complex
  organic matter in Titan's atmospheric aerosols from in situ pyrolysis and
  analysis}. Nature 438, 796--799.

\bibitem[{{Isra{\"e}l} et~al.(2006){Isra{\"e}l}, {Szopa}, {Raulin}, {Cabane},
  {Niemann}, {Atreya}, {Bauer}, {Brun}, {Chassefi{\`e}re}, {Coll}, {Cond{\'e}},
  {Coscia}, {Hauchecorne}, {Millian}, {Nguyen}, {Owen}, {Riedler}, {Samuelson},
  {Siguier}, {Steller}, {Sternberg}, and {Vidal-Madjar}}]{Israel:2006}
{Isra{\"e}l}, G., {Szopa}, C., {Raulin}, F., {Cabane}, M., {Niemann}, H.~B.,
  {Atreya}, S.~K., {Bauer}, S.~J., {Brun}, J.-F., {Chassefi{\`e}re}, E.,
  {Coll}, P., {Cond{\'e}}, E., {Coscia}, D., {Hauchecorne}, A., {Millian}, P.,
  {Nguyen}, M.~J., {Owen}, T., {Riedler}, W., {Samuelson}, R.~E., {Siguier},
  J.-M., {Steller}, M., {Sternberg}, R., {Vidal-Madjar}, C., Nov. 2006.
  {Astrochemistry: Complex organic matter in Titan's aerosols? (Reply)}. Nature
  444.

\bibitem[{Jimenez et~al.({2003}{\natexlab{a}})Jimenez, Bahreini, Cocker,
  Zhuang, Varutbangkul, Flagan, Seinfeld, O'Dowd, and Hoffmann}]{Jimenez:2003a}
Jimenez, J., Bahreini, R., Cocker, D., Zhuang, H., Varutbangkul, V., Flagan,
  R., Seinfeld, J., O'Dowd, C., Hoffmann, T., {MAY 30} {2003}{\natexlab{a}}.
  {New particle formation from photooxidation of diiodomethane (CH2I2)}.
  {Journal of Geophysical Research (Atmospheres)} {108}~({D10}).

\bibitem[{Jimenez et~al.({2003}{\natexlab{b}})Jimenez, Bahreini, Cocker,
  Zhuang, Varutbangkul, Flagan, Seinfeld, O'Dowd, and Hoffmann}]{Jimenez:2003b}
Jimenez, J., Bahreini, R., Cocker, D., Zhuang, H., Varutbangkul, V., Flagan,
  R., Seinfeld, J., O'Dowd, C., Hoffmann, T., {DEC 10} {2003}{\natexlab{b}}.
  {New particle formation from photooxidation of diiodomethane (CH2I2) (vol
  108, Art no 4318, 2003)}. {Journal of Geophysical Research (Atmospheres)}
  {108}~({D23}).

\bibitem[{Kameta et~al.({2002})Kameta, Kouchi, Ukai, and Hatano}]{Kameta:2002}
Kameta, K., Kouchi, N., Ukai, M., Hatano, Y., {MAY} {2002}. {Photoabsorption,
  photoionization, and neutral-dissociation cross sections of simple
  hydrocarbons in the vacuum ultraviolet range}. {Journal of Electron
  Spectroscopy and Related Phenomena} {123}~({2-3}), {225--238}.

\bibitem[{{Khare} et~al.(1984){Khare}, {Sagan}, {Thompson}, {Arakawa}, {Suits},
  {Callcott}, {Williams}, {Shrader}, {Ogino}, {Willingham}, and
  {Nagy}}]{Khare:1984}
{Khare}, B.~N., {Sagan}, C., {Thompson}, W.~R., {Arakawa}, E.~T., {Suits}, F.,
  {Callcott}, T.~A., {Williams}, M.~W., {Shrader}, S., {Ogino}, H.,
  {Willingham}, T.~O., {Nagy}, B., 1984. {The organic aerosols of Titan}. Adv.
  Space Res. 4, 59--68.

\bibitem[{Kroll et~al.({2011})Kroll, Donahue, Jimenez, Kessler, Canagaratna,
  Wilson, Altieri, Mazzoleni, Wozniak, Bluhm, Mysak, Smith, Kolb, and
  Worsnop}]{Kroll:2011}
Kroll, J.~H., Donahue, N.~M., Jimenez, J.~L., Kessler, S.~H., Canagaratna,
  M.~R., Wilson, K.~R., Altieri, K.~E., Mazzoleni, L.~R., Wozniak, A.~S.,
  Bluhm, H., Mysak, E.~R., Smith, J.~D., Kolb, C.~E., Worsnop, D.~R., {FEB}
  {2011}. {Carbon oxidation state as a metric for describing the chemistry of
  atmospheric organic aerosol}. {Nature Chemistry} {3}~({2}), {133--139}.

\bibitem[{{Lavvas} et~al.(2011){Lavvas}, {Sander}, {Kraft}, and
  {Imanaka}}]{Lavvas:2011}
{Lavvas}, P., {Sander}, M., {Kraft}, M., {Imanaka}, H., Feb. 2011. {Surface
  Chemistry and Particle Shape: Processes for the Evolution of Aerosols in
  Titan's Atmosphere}. ApJ 728, 80--+.

\bibitem[{{Liang} et~al.(2007){Liang}, {Yung}, and {Shemansky}}]{Liang:2007}
{Liang}, M., {Yung}, Y.~L., {Shemansky}, D.~E., Jun. 2007. {Photolytically
  Generated Aerosols in the Mesosphere and Thermosphere of Titan}. ApJL 661,
  L199--L202.

\bibitem[{{Magee} et~al.(2009){Magee}, {Waite}, {Mandt}, {Westlake}, {Bell},
  and {Gell}}]{Magee:2009}
{Magee}, B.~A., {Waite}, J.~H., {Mandt}, K.~E., {Westlake}, J., {Bell}, J.,
  {Gell}, D.~A., Dec. 2009. {INMS-derived composition of Titan's upper
  atmosphere: Analysis methods and model comparison}. Planet. Space Sci. 57,
  1895--1916.

\bibitem[{{Marten} et~al.(2002){Marten}, {Hidayat}, {Biraud}, and
  {Moreno}}]{Marten:2002}
{Marten}, A., {Hidayat}, T., {Biraud}, Y., {Moreno}, R., Aug. 2002. {New
  Millimeter Heterodyne Observations of Titan: Vertical Distributions of
  Nitriles HCN, HC$_{3}$N, CH$_{3}$CN, and the Isotopic Ratio $^{15}$N/$^{14}$N
  in Its Atmosphere}. Icarus 158, 532--544.

\bibitem[{{McDonald} et~al.(1994){McDonald}, {Thompson}, {Heinrich}, {Khare},
  and {Sagan}}]{McDonald:1994}
{McDonald}, G.~D., {Thompson}, W.~R., {Heinrich}, M., {Khare}, B.~N., {Sagan},
  C., Mar. 1994. {Chemical investigation of Titan and Triton tholins}. Icarus
  108, 137--145.

\bibitem[{{McLafferty} and Ture\v{c}ek(1993)}]{Mclafferty:1993}
{McLafferty}, F.~W., Ture\v{c}ek, F., 1993. {Interpretation of Mass Spectra}.
  University Science Books, Sausalito.

\bibitem[{{Mount} et~al.(1977){Mount}, {Warden}, and {Moos}}]{Mount:1977}
{Mount}, G.~H., {Warden}, E.~S., {Moos}, H.~W., May 1977. {Photoabsorption
  cross sections of methane from 1400 to 1850 A}. ApJL 214, L47--L49.

\bibitem[{{Navarro-Gonz{\'a}lez} and {Ram{\'{\i}}rez}(1997)}]{Navarro:1997}
{Navarro-Gonz{\'a}lez}, R., {Ram{\'{\i}}rez}, S.~I., May 1997. {Corona
  discharge of Titan's troposphere}. Advances in Space Research 19, 1121--1133.

\bibitem[{{Nixon} et~al.(2013){Nixon}, {Jennings}, {B{\'e}zard}, {Vinatier},
  {Teanby}, {Sung}, {Ansty}, {Irwin}, {Gorius}, {Cottini}, {Coustenis}, and
  {Flasar}}]{Nixon:2013b}
{Nixon}, C.~A., {Jennings}, D.~E., {B{\'e}zard}, B., {Vinatier}, S., {Teanby},
  N.~A., {Sung}, K., {Ansty}, T.~M., {Irwin}, P.~G.~J., {Gorius}, N.,
  {Cottini}, V., {Coustenis}, A., {Flasar}, F.~M., Oct. 2013. {Detection of
  Propene in Titan's Stratosphere}. ApJL 776, L14.

\bibitem[{{Peng} et~al.(2013){Peng}, {Gautier}, {Carrasco}, {Pernot},
  {Giuliani}, {Mahjoub}, {Correia}, {Buch}, {B{\'e}nilan}, {Szopa}, and
  {Cernogora}}]{Peng:2013}
{Peng}, Z., {Gautier}, T., {Carrasco}, N., {Pernot}, P., {Giuliani}, A.,
  {Mahjoub}, A., {Correia}, J.-J., {Buch}, A., {B{\'e}nilan}, Y., {Szopa}, C.,
  {Cernogora}, G., Apr. 2013. {Titan's atmosphere simulation experiment using
  continuum UV-VUV synchrotron radiation}. Journal of Geophysical Research
  (Planets) 118, 778--788.

\bibitem[{Rajappan et~al.(2010)Rajappan, B殳tner, Cox, and
  Yates~Jr}]{Rajappan:2010}
Rajappan, M., B殳tner, M., Cox, C., Yates~Jr, J.~T., 2010. Photochemical
  decomposition of n2o by lyman-$\alpha$ radiation: Scientific basis for a
  chemical actinometer. The Journal of Physical Chemistry A 114~(10),
  3443--3448.

\bibitem[{{Ram{\'{\i}}rez} et~al.(2001){Ram{\'{\i}}rez},
  {Navarro-Gonz{\'a}lez}, {Coll}, and {Raulin}}]{Ramirez:2001}
{Ram{\'{\i}}rez}, S.~I., {Navarro-Gonz{\'a}lez}, R., {Coll}, P., {Raulin}, F.,
  2001. {Possible contribution of different energy sources to the production of
  organics in Titan's atmosphere}. Adv. Space Res. 27, 261--270.

\bibitem[{{Ram{\'{\i}}rez} et~al.(2005){Ram{\'{\i}}rez},
  {Navarro-Gonz{\'a}lez}, {Coll}, and {Raulin}}]{Ramirez:2005}
{Ram{\'{\i}}rez}, S.~I., {Navarro-Gonz{\'a}lez}, R., {Coll}, P., {Raulin}, F.,
  2005. {Organic chemistry induced by corona discharges in Titan's troposphere:
  Laboratory simulations}. Adv. Space Res. 36, 274--280.

\bibitem[{{Raulin} et~al.(1982){Raulin}, {Mourey}, and
  {Toupance}}]{Raulin:1982}
{Raulin}, F., {Mourey}, D., {Toupance}, G., Sep. 1982. {Organic syntheses from
  CH$_{4}$-N$_{2}$ atmospheres: Implications for Titan}. Origins of Life 12,
  267--279.

\bibitem[{{Sarker} et~al.(2003){Sarker}, {Somogyi}, {Lunine}, and
  {Smith}}]{Sarker:2003}
{Sarker}, N., {Somogyi}, A., {Lunine}, J.~I., {Smith}, M.~A., Dec. 2003. {Titan
  Aerosol Analogues: Analysis of the Nonvolatile Tholins}. Astrobiology 3,
  719--726.

\bibitem[{{Scattergood} and {Owen}(1977)}]{Scattergood:1977}
{Scattergood}, T., {Owen}, T., Apr. 1977. {On the sources of ultraviolet
  absorption in spectra of Titan and the outer planets}. Icarus 30, 780--788.

\bibitem[{{Scattergood} et~al.(1989){Scattergood}, {McKay}, {Borucki}, {Giver},
  {van Ghyseghem}, {Parris}, and {Miller}}]{Scattergood:1989}
{Scattergood}, T.~W., {McKay}, C.~P., {Borucki}, W.~J., {Giver}, L.~P., {van
  Ghyseghem}, H., {Parris}, J.~E., {Miller}, S.~L., Oct. 1989. {Production of
  organic compounds in plasmas - A comparison among electric sparks,
  laser-induced plasmas, and UV light}. Icarus 81, 413--428.

\bibitem[{{Sciamma-O'Brien} et~al.(2010){Sciamma-O'Brien}, {Carrasco}, {Szopa},
  {Buch}, and {Cernogora}}]{Sciamma:2010}
{Sciamma-O'Brien}, E., {Carrasco}, N., {Szopa}, C., {Buch}, A., {Cernogora},
  G., Oct. 2010. {Titan's atmosphere: An optimal gas mixture for aerosol
  production?} Icarus 209, 704--714.

\bibitem[{{Sciamma-O'Brien} et~al.(2014){Sciamma-O'Brien}, {Ricketts}, and
  {Salama}}]{Sciamma:2014}
{Sciamma-O'Brien}, E., {Ricketts}, C.~L., {Salama}, F., Nov. 2014. {The Titan
  Haze Simulation experiment on COSmIC: Probing Titan's atmospheric chemistry
  at low temperature}. Icarus 243, 325--336.

\bibitem[{{Sebree} et~al.(2014){Sebree}, {Trainer}, {Loeffler}, and
  {Anderson}}]{Sebree:2014}
{Sebree}, J.~A., {Trainer}, M.~G., {Loeffler}, M.~J., {Anderson}, C.~M., Jul.
  2014. {Titan aerosol analog absorption features produced from aromatics in
  the far infrared}. Icarus 236, 146--152.

\bibitem[{{Sekine} et~al.(2008){Sekine}, {Imanaka}, {Matsui}, {Khare}, {Bakes},
  {McKay}, and {Sugita}}]{Sekine:2008}
{Sekine}, Y., {Imanaka}, H., {Matsui}, T., {Khare}, B.~N., {Bakes}, E.~L.~O.,
  {McKay}, C.~P., {Sugita}, S., Mar. 2008. {The role of organic haze in Titan's
  atmospheric chemistry. I. Laboratory investigation on heterogeneous reaction
  of atomic hydrogen with Titan tholin}. Icarus 194, 186--200.

\bibitem[{{Teanby} et~al.(2013){Teanby}, {Irwin}, {Nixon}, {Courtin},
  {Swinyard}, {Moreno}, {Lellouch}, {Rengel}, and {Hartogh}}]{Teanby:2013}
{Teanby}, N.~A., {Irwin}, P.~G.~J., {Nixon}, C.~A., {Courtin}, R., {Swinyard},
  B.~M., {Moreno}, R., {Lellouch}, E., {Rengel}, M., {Hartogh}, P., Jan. 2013.
  {Constraints on Titan's middle atmosphere ammonia abundance from
  Herschel/SPIRE sub-millimetre spectra}. Planet. Space Sci. 75, 136--147.

\bibitem[{{Thompson} et~al.(1991){Thompson}, {Henry}, {Schwartz}, {Khare}, and
  {Sagan}}]{Thompson:1991}
{Thompson}, W.~R., {Henry}, T.~J., {Schwartz}, J.~M., {Khare}, B.~N., {Sagan},
  C., Mar. 1991. {Plasma discharge in N$_{2}$ + CH$_{4}$ at low pressures -
  Experimental results and applications to Titan}. Icarus 90, 57--73.

\bibitem[{{Trainer} et~al.(2012){Trainer}, {Jimenez}, {Yung}, {Toon}, and
  {Tolbert}}]{Trainer:2012}
{Trainer}, M.~G., {Jimenez}, J.~L., {Yung}, Y.~L., {Toon}, O.~B., {Tolbert},
  M.~A., Apr. 2012. {Nitrogen Incorporation in CH$_{4}$-N$_{2}$ Photochemical
  Aerosol Produced by Far Ultraviolet Irradiation}. Astrobiology 12, 315--326.

\bibitem[{{Trainer} et~al.(2004{\natexlab{a}}){Trainer}, {Pavlov}, {Curtis},
  {McKay}, {Worsnop}, {Delia}, {Toohey}, {Toon}, and {Tolbert}}]{Trainer:2004b}
{Trainer}, M.~G., {Pavlov}, A.~A., {Curtis}, D.~B., {McKay}, C.~P., {Worsnop},
  D.~R., {Delia}, A.~E., {Toohey}, D.~W., {Toon}, O.~B., {Tolbert}, M.~A., Dec.
  2004{\natexlab{a}}. {Haze Aerosols in the Atmosphere of Early Earth: Manna
  from Heaven}. Astrobiology 4, 409--419.

\bibitem[{{Trainer} et~al.(2006){Trainer}, {Pavlov}, {DeWitt}, {Jimenez},
  {McKay}, {Toon}, and {Tolbert}}]{Trainer:2006}
{Trainer}, M.~G., {Pavlov}, A.~A., {DeWitt}, H.~L., {Jimenez}, J.~L., {McKay},
  C.~P., {Toon}, O.~B., {Tolbert}, M.~A., Nov. 2006. {Organic Haze on Titan and
  the early Earth}. PNAS 103, 18035--18042.

\bibitem[{{Trainer} et~al.(2004{\natexlab{b}}){Trainer}, {Pavlov}, {Jimenez},
  {McKay}, {Worsnop}, {Toon}, and {Tolbert}}]{Trainer:2004}
{Trainer}, M.~G., {Pavlov}, A.~A., {Jimenez}, J.~L., {McKay}, C.~P., {Worsnop},
  D.~R., {Toon}, O.~B., {Tolbert}, M.~A., Aug. 2004{\natexlab{b}}. {Chemical
  composition of Titan's haze: Are PAHs present?} Geophys. Res. Lett. 31,
  17--+.

\bibitem[{{Trainer} et~al.(2013){Trainer}, {Sebree}, {Yoon}, and
  {Tolbert}}]{Trainer:2013}
{Trainer}, M.~G., {Sebree}, J.~A., {Yoon}, Y.~H., {Tolbert}, M.~A., Mar. 2013.
  {The Influence of Benzene as a Trace Reactant in Titan Aerosol Analogs}. ApJL
  766, L4.

\bibitem[{{Tran} et~al.(2003{\natexlab{a}}){Tran}, {Ferris}, and
  {Chera}}]{Tran:2003}
{Tran}, B.~N., {Ferris}, J.~P., {Chera}, J.~J., Mar. 2003{\natexlab{a}}. {The
  photochemical formation of a titan haze analog. Structural analysis by x-ray
  photoelectron and infrared spectroscopy}. Icarus 162, 114--124.

\bibitem[{{Tran} et~al.(2008){Tran}, {Force}, {Briggs}, {Ferris}, {Persans},
  and {Chera}}]{Tran:2008}
{Tran}, B.~N., {Force}, M., {Briggs}, R.~G., {Ferris}, J.~P., {Persans}, P.,
  {Chera}, J.~J., Jan. 2008. {Titan's atmospheric chemistry: Photolysis of gas
  mixtures containing hydrogen cyanide and carbon monoxide at 185 and 254 nm}.
  Icarus 193, 224--232.

\bibitem[{{Tran} et~al.(2003{\natexlab{b}}){Tran}, {Joseph}, {Ferris},
  {Persans}, and {Chera}}]{Tran:2003b}
{Tran}, B.~N., {Joseph}, J.~C., {Ferris}, J.~P., {Persans}, P.~D., {Chera},
  J.~J., Oct. 2003{\natexlab{b}}. {Simulation of Titan haze formation using a
  photochemical flow reactor The optical constants of the polymer}. Icarus 165,
  379--390.

\bibitem[{{Vuitton} et~al.(2007){Vuitton}, {Yelle}, and
  {McEwan}}]{Vuitton:2007}
{Vuitton}, V., {Yelle}, R.~V., {McEwan}, M.~J., Nov. 2007. {Ion chemistry and
  N-containing molecules in Titan's upper atmosphere}. Icarus 191, 722--742.

\bibitem[{Warneke et~al.({2005})Warneke, de~Gouw, Lovejoy, Murphy, Kuster, and
  Fall}]{Warneke:2005}
Warneke, C., de~Gouw, J., Lovejoy, E., Murphy, P., Kuster, W., Fall, R., {AUG}
  {2005}. {Development of proton-transfer ion trap-mass spectrometry: On-line
  detection and identification of volatile organic compounds in air}. {Journal
  of the American Society of Mass Spectrometry} {16}~({8}), {1316--1324}.

\bibitem[{Warneke et~al.({2011})Warneke, Roberts, Veres, Gilman, Kuster,
  Burling, Yokelson, and de~Gouw}]{Warneke:2011}
Warneke, C., Roberts, J.~M., Veres, P., Gilman, J., Kuster, W.~C., Burling, I.,
  Yokelson, R., de~Gouw, J.~A., {MAY 15} {2011}. {VOC identification and
  inter-comparison from laboratory biomass burning using PTR-MS and PIT-MS}.
  {International Journal of Mass Spectrometry} {303}~({1}), {6--14}.

\bibitem[{{Warneke} et~al.(2011){Warneke}, {Veres}, {Holloway}, {Stutz},
  {Tsai}, {Alvarez}, {Rappenglueck}, {Fehsenfeld}, {Graus}, {Gilman}, and {de
  Gouw}}]{Warneke:2011b}
{Warneke}, C., {Veres}, P., {Holloway}, J.~S., {Stutz}, J., {Tsai}, C.,
  {Alvarez}, S., {Rappenglueck}, B., {Fehsenfeld}, F.~C., {Graus}, M.,
  {Gilman}, J.~B., {de Gouw}, J.~A., Oct. 2011. {Airborne formaldehyde
  measurements using PTR-MS: calibration, humidity dependence, inter-comparison
  and initial results}. Atmospheric Measurement Techniques 4, 2345--2358.

\bibitem[{{Watson} and {Sparkman}(2007)}]{Watson:2007}
{Watson}, J.~T., {Sparkman}, D.~O., 2007. {Introduction to Mass Spectrometry}.
  Wiley, Orlando.

\bibitem[{{Woods} et~al.(2009){Woods}, {Chamberlin}, {Harder}, {Hock}, {Snow},
  {Eparvier}, {Fontenla}, {McClintock}, and {Richard}}]{Woods:2009}
{Woods}, T.~N., {Chamberlin}, P.~C., {Harder}, J.~W., {Hock}, R.~A., {Snow},
  M., {Eparvier}, F.~G., {Fontenla}, J., {McClintock}, W.~E., {Richard}, E.~C.,
  Jan. 2009. {Solar Irradiance Reference Spectra (SIRS) for the 2008 Whole
  Heliosphere Interval (WHI)}. Geophys. Res. Lett. 36, 1101.

\bibitem[{Yelle et~al.(2010)Yelle, Vuitton, Lavvas, Klippenstein, Smith,
  H{\"o}rst, and Cui}]{Yelle:2010}
Yelle, R.~V., Vuitton, V., Lavvas, P., Klippenstein, S.~J., Smith, M.~A.,
  H{\"o}rst, S.~M., Cui, J., 2010. {Formation of NH$_{3}$ and CH$_{2}$NH in
  Titan's upper atmosphere}. Faraday Discuss. 147, 31--49.

\bibitem[{{Yoon} et~al.(2014){Yoon}, {H{\"o}rst}, {Hicks}, {Li}, {de Gouw}, and
  {Tolbert}}]{Yoon:2014}
{Yoon}, Y.~H., {H{\"o}rst}, S.~M., {Hicks}, R.~K., {Li}, R., {de Gouw}, J.~A.,
  {Tolbert}, M.~A., May 2014. {The role of benzene photolysis in Titan haze
  formation}. Icarus 233, 233--241.

\end{thebibliography}







\end{document}